\documentclass[11pt,draftclsnofoot,onecolumn]{IEEEtran}
\usepackage{cite}
\usepackage{graphicx}
\usepackage{psfrag}
\usepackage{subfigure}
\usepackage{url}
\usepackage{stfloats}
\usepackage{amsmath}
\usepackage{array}
\usepackage{amsmath}
\usepackage{amsfonts}
\usepackage{amssymb}
\usepackage{epsfig}
\usepackage{algorithm}
\usepackage{mathrsfs}
\usepackage{algpseudocode}
\usepackage{dsfont}
\usepackage{multirow}

\newtheorem{theorem}{{Theorem}}
\newtheorem{lemma}{{Lemma}}
\newtheorem{definition}{{Definition}}
\newtheorem{remark}{{Remark}}
\newtheorem{corollary}{{Corollary}}
\newtheorem{example}{{Example}}

\begin{document}
\title{Structural Solutions For Additively\\Coupled Sum Constrained Games}
\author{Yi~Su and~Mihaela van der Schaar
\\ Department of Electrical Engineering, UCLA}
\maketitle

\begin{abstract}
We propose and analyze a broad family of games played by
resource-constrained players, which are characterized by the
following central features: 1) each user has a multi-dimensional
action space, subject to a single sum resource constraint; 2) each
user's utility in a particular dimension depends on an additive
coupling between the user's action in the same dimension and the
actions of the other users; and 3) each user's total utility is the
sum of the utilities obtained in each dimension. Familiar examples
of such multi-user environments in communication systems include
power control over frequency-selective Gaussian interference
channels and flow control in Jackson networks. In settings where
users cannot exchange messages in real-time, we study how users can
adjust their actions based on their local observations. We derive
sufficient conditions under which a unique Nash equilibrium exists
and the best-response algorithm converges globally and linearly to
the Nash equilibrium. In settings where users can exchange messages
in real-time, we focus on user choices that optimize the overall
utility. We provide the convergence conditions of two distributed
action update mechanisms, gradient play and Jacobi update.
\end{abstract}

\begin{IEEEkeywords}
Game theory, multi-user communications, Nash equilibrium,
best-response dynamics, Gradient play, Jacobi update, pricing
mechanism.
\end{IEEEkeywords}
\IEEEpeerreviewmaketitle

\section{Introduction}
Game theory provides a formal framework for describing and analyzing
the interactions of multiple decision-makers. Recently, there has
been a surge in research activities that adopt game theoretic tools
to investigate a wide range of modern communications and networking
problems, such as flow and congestion control, network routing, load
balancing, power control, peer-to-peer content sharing, etc
\cite{Survery_Altman}-\cite{SPM_GT}. In resource-constrained
communication networks, a user's utility is usually not only
affected by its own action but also by the actions taken by all the
other users sharing the same resources. Due to the mutual coupling
among users, the performance optimization of multi-user
communication systems is challenging. Depending on the
characteristics of different applications, numerous game-theoretical
models and solution concepts have been proposed to characterize the
multi-user interactions and optimize the users' decisions in
communication networks. A variety of game theoretic solutions have
been developed to characterize the resulting performance of the
multi-user interactions, including Nash equilibrium (NE) and Pareto
optimality \cite{Game_book}.

The majority of the existing game theoretic research works in
communication networking applications usually depend on the specific
structures and inter-user coupling of their action sets and utility
functions. By considering or even architecting these specific
structures, the associated games become analytically tractable and
possess various important convergence properties. For instance, if
users cannot exchange messages with each other and choose to
individually maximize their utilities, to show the existence of and
the convergence to a pure NE, several well-investigated classes of
game models, such as concave games, supermodular games, and
potential games, have been extensively applied in various
communication scenarios \cite{Game_book}-\cite{Pgame_1}. When
real-time information exchange is possible, various mechanisms have
also been proposed to enable collaborative users to jointly improve
their performance and find the optimum joint policy. A well-known
example is the framework of network utility maximization (NUM)
started by Kelly etc. \cite{Kelly}\cite{Num}, which has recently
been widely adopted to analyze the problems related to fairness and
efficiency in communication networks. Moreover, various distributed
resource allocation algorithms have been developed to implement the
NUM framework in an informationally-decentralized manner. In
particular, if a convex NUM problem can be decomposed into several
subproblems by introducing Lagrange multipliers associated with
different resource constraints, the global optimum can be computed
using distributed algorithms by deploying message passing mechanisms
\cite{Daniel_tutorial}.

Power control is one of the first few communication problems in
which researchers started to apply game theoretic tools to formalize
the multi-user interaction and characterize its properties. An
interesting and important topic that has been extensively
investigated recently is how to optimize multiple devices' power
allocation when sharing a common frequency-selective interference
channel. In \cite{Yu_JSAC}, Yu et. al. first defined such a power
control game from a game-theoretic perspective, proposed a
best-response algorithm in which all users iteratively update their
power allocations using the water-filling solution, and proved
several sufficient conditions under which the algorithm globally
converges to a unique pure NE. Many follow-up papers further
establish various sufficient convergence conditions with or without
real-time information exchange for power control in communication
networks \cite{Chung}-\cite{Shi}. The purpose of this paper is to
introduce and analyze a general framework that abstracts the common
characteristics of this family of multi-user interaction scenarios,
which includes, but is not limited to, the power control scenario.
In particular, the main contributions of this paper are as follows.

First of all, we define the class of \emph{Additively Coupled Sum
Constrained Games} (ACSCG), which captures and characterizes the key
features of several communication and networking applications. In
particular, the central features of ACSCG are: 1) each user has a
multi-dimensional strategy that is subject to a single sum resource
constraint; 2) each user's payoff in each dimension is impacted by
an additive combination of its own action in the same dimension and
a function of the other users' actions; 3) users' utilities are
separable across different dimensions and each user's total utility
is the sum of the utilities obtained within each dimension.

Second, based on the feasibility of real-time information exchange,
we provide the convergence conditions of various generic distributed
algorithms in different scenarios. When no message exchanges between
users are possible and every user maximizes its own utility, it is
essential to determine whether a NE exist and if yes, how to achieve
such an equilibrium. In ACSCG, a pure NE exists in ACSCG because
ACSCG belongs to concave games \cite{Game_book}\cite{Rosen}. Our key
contribution in this context is that we investigate the uniqueness
of pure NE and consider the best response dynamics to compute the
NE. We explore the properties of the additive coupling among users
given the sum constraint and provide several sufficient conditions
under which best response dynamics converges linearly\footnote{A
sequence $x^{(k)}$ with limit $x^*$ is linearly convergent if there
exists a constant $c\in(0,1)$ such that $|x^{(k)}-x^*|\leq
c|x^{(k-1)}-x^*|$ for $k$ sufficiently large \cite{convex_book}.} to
the unique NE, for any set of feasible initialization with either
sequential or parallel updates. We also explain the relationship
between our results and the conditions previously developed in the
game theory literature \cite{Rosen}\cite{Moulin}. When users can
collaboratively exchange messages with each other in real-time, we
present the sufficient convergence conditions of two alternative
distributed pricing algorithms, including gradient play and Jacobi
update, to coordinate users' action and improve the overall system
efficiency. The proposed convergence conditions generalize the
results that have been previously obtained in
\cite{Yu_JSAC}-\cite{Shi} for the multi-user power control problem
and they are immediately applicable to other multi-user applications
in communication networks that fulfill the requirements of ACSCG.

The rest of this paper is organized as follows. Section II defines
the model of ACSCG. For ACSCG models, Sections III and IV present
several distributed algorithms without and with real-time
information exchanges, respectively, and provide sufficient
conditions that guarantee the convergence of the proposed
algorithms. Section V presents the numerical examples and
conclusions are drawn in Section VI.

\section{Game Model}
In this section, we introduce some basic definitions from the theory
of strategic games to characterize the multi-user interaction,
define the model of ACSCG, and present some illustrative examples
for the class of ACSCG.

\subsection{Strategic Games, Nash equilibrium, and Pareto Optimality}
A strategic game is a suitable model for the analysis of a game
where all users act independently and simultaneously according to
their own self-interests and with no or limited a priori knowledge
of the other users' strategies. This can be formally defined as a
tuple $\Gamma=\langle\mathcal{N},\mathcal{A},u\rangle$. In
particular, $\mathcal{N}=\{1,2,\ldots,N\}$ is the set of decision
makers. Define $\mathcal{A}$ to be the joint action set
$\mathcal{A}=\times_{n\in\mathcal{N}} \mathcal{A}_n$, with
$\mathcal{A}_n\subseteq \mathcal{R}^K$ being the action set
available for user $n$. The vector utility function
$u=\times_{n\in\mathcal{N}}u_n$ is a mapping from the individual
users' joint action set to real numbers, i.e. $u : \mathcal{A}
\rightarrow \mathcal{R}^N$. In particular, $u_n(\textbf{a}) :
\mathcal{A} \rightarrow \mathcal{R}$ is the utility of the $n$th
user that generally depends on the strategies
$\textbf{a}=(\textbf{a}_n, \textbf{a}_{-n})$ of all users, where
$\textbf{a}_n \in \mathcal{A}_n$ denotes a feasible action of user
$n$, and $\textbf{a}_{-n}=\times_{m\neq n}\textbf{a}_m$ is a vector
of the actions of all users except $n$. We also denote by
$\mathcal{A}_{-n}=\times_{m\neq n}\mathcal{A}_m$ the joint action
set of all users except $n$. To capture the multi-user performance
tradeoff, the utility region is defined as
$\mathcal{U}=\{(u_1(\textbf{a}),\ldots,u_N(\textbf{a}))| \ \exists \
\textbf{a}\in \mathcal{A}\}$. Various game theoretic solutions, such
as NE and Pareto optimality, were developed to characterize the
resulting performance\cite{Game_book}. Significant research efforts
have been devoted in the literature to constructing operational
algorithms in order to achieve NE and Pareto optimality in various
games with special structures of action set $\mathcal{A}_n$ and
utility function $u_n$.

\subsubsection{Nash equilibrium: definition, existence, and convergence}
To avoid the overhead associated with exchanging information in
real-time, network designers may prefer fully decentralized
solutions in which the participating users simply compete against
other users by choosing actions $\textbf{a}_n \in \mathcal{A}_n$ to
selfishly maximize their individual utility functions
$u_n(\textbf{a}_n, \textbf{a}_{-n})$, given the actions
$\textbf{a}_{-n} \in \mathcal{A}_{-n}$. Most of these approaches
focus on investigating the existence and properties of NE. NE is
defined to be an action profile
$(\textbf{a}_1^*,\textbf{a}_2^*,\ldots,\textbf{a}_N^*)$ with the
property that for every player, it satisfies $u_n(\textbf{a}_n^*,
\textbf{a}_{-n}^*)\geq u_n(\textbf{a}_n, \textbf{a}_{-n}^*)$ for all
$\textbf{a}_n\in \mathcal{A}_n$, i.e. given the other users'
actions, no user can increase its utility alone by changing its
action. For an extensive discussion of the methodologies studying
the existence, uniqueness, and convergence of various equilibria in
communication networks, we refer the readers to \cite{Altman_SPM}.
Many of the well-known results rely on specific structural
properties of action set $\mathcal{A}$ and utility function $u$ in
the investigated multi-user interactions. For example, to establish
the existence of and convergence to a pure NE, we can examine
whether $\mathcal{A}$ and $u$ satisfy the conditions of concave
games, supermodular game, potential game, etc. Specifically, to
apply the existence result of a pure NE in concave games
\cite{Game_book}\cite{Rosen}, we need to check the following
conditions: i) each player's action set $\mathcal{A}_n$ is convex
and compact; and ii) the utility function $u_n(\textbf{a}_n,
\textbf{a}_{-n})$ is continuous in $\textbf{a}$ and
quasi-concave\footnote{A real-valued function $f$ is quasi-concave
if $\textmd{dom}f$ is convex and
$\{x\in\textmd{dom}f|f(x)\geq\alpha\}$ is convex for all $\alpha$.}
in $\textbf{a}_n$ for any fixed $\textbf{a}_{-n}$. As additional
examples of games that guarantee the convergence to NE, it is
well-known that, in supermodular games
\cite{Topkis_book}\cite{Sgame_2} and potential games
\cite{Pgame_Shapley}\cite{Pgame_1}, the best response dynamics can
be used to search for a pure NE. Suppose that utility function $u_n$
is twice continuously differentiable, $\forall n \in \mathcal{N}$.
If $\mathcal{A}_n$ is a compact subset of $\mathcal{R}$ (or more
generally $\mathcal{A}_n$ is a nonempty and compact
sublattice\footnote{A real $K$-dimensional set $\mathcal{V}$ is a
\textit{sublattice} of $\mathcal{R}^K$ if for any two elements $a,b
\in \mathcal{V}$, the component-wise minimum, $a\wedge b$, and the
component-wise maximum, $a\vee b$, are also in $\mathcal{V}$.} of
$\mathcal{R}^K$), $\forall n \in \mathcal{N}$, establishing that
game $\Gamma$ is a supermodular game is equivalent to showing that
$u_n$ satisfies
\begin{equation}
\label{eq:eqn1} \forall (m,n) \in \mathcal{N}^2, m \neq n,
\frac{\partial^2 u_n}{\partial \textbf{a}_n\partial
\textbf{a}_m}\geq 0.
\end{equation} If action set
$\mathcal{A}$ in game $\Gamma$ is an interval of real numbers, we
can show that game $\Gamma$ is a potential game by verifying
\begin{equation}
\forall (m,n) \in \mathcal{N}^2, m \neq n,
\frac{\partial^2(u_n-u_m)}{\partial \textbf{a}_n\partial
\textbf{a}_m}=0.
\end{equation}

\subsubsection{Pareto optimality and network utility maximization}
It is important to note that operating at a Nash equilibrium will
generally limit the performance of the user itself as well as that
of the entire network, because the available network resources are
not always effectively exploited due to the conflicts of interest
occurring among users. As opposed to the NE-based approaches, there
exists a large body of literature that focuses on studying how users
can \textit{jointly} improve the system performance by optimizing a
certain common objective function
$f(u_1(\textbf{a}),u_2(\textbf{a}),\ldots,u_N(\textbf{a}))$. This
function represents the fairness rule based on which the system-wide
resource allocation is performed. Different objective functions,
e.g. sum utility maximization in which
$f(u_1(\textbf{a}),u_2(\textbf{a}),\ldots,u_N(\textbf{a}))=\sum_{n=1}^N
u_n(\textbf{a})$, can provide reasonable allocation outcomes by
jointly considering fairness and efficiency. A profile of actions is
Pareto optimal if there is no other profile of actions that makes
every user at least as well off and at least one user strictly
better off.

The majority of these approaches focus on studying how to
efficiently or distributedly find the optimum joint policy. There
exists a large body of literature that investigates how to compute
Pareto optimal solutions in large-scale networks where centralized
solutions are infeasible. Numerous convergence results have been
obtained for various generic distributed algorithms. An important
example is the NUM framework that develops distributed algorithms to
solve network resource allocation problems \cite{Num}. The majority
of the results in the existing NUM literature are based on convex
optimization theory, in which the investigated problems share the
following structures: the objective function
$f(u_1(\textbf{a}),u_2(\textbf{a}),\ldots,u_N(\textbf{a}))$ is
convex\footnote{$f:\textmd{R}^n\rightarrow \textmd{R}$ is convex if
$\textmd{dom}f$ is a convex set and $f(\theta x+(1-\theta)y)\leq
\theta f(x)+(1-\theta)f(y)$, $\forall x,y \in \textmd{dom}f,
0\leq\theta\leq1$.}, inequality resource constraint functions are
convex, and equality resource constraint functions are affine. It is
well-known that, for convex optimization problems, users can
collaboratively exchange price signals that reflect the ``cost" for
consuming the constrained resources and the Pareto optimal
allocation that maximizes the network utility can be determined in a
fully distributed manner \cite{Daniel_tutorial}.

Summarizing, these general structural results without and with
real-time message exchange turn out to be very useful when analyzing
various multi-user interactions in communication networks. Numerous
existing works are devoted to constructing or shaping the multi-user
coupling such that it fits into these frameworks and the
corresponding generic solutions can be directly applied. In the
remaining part of this paper, we will derive several structural
results for a particular type of multi-user interaction scenario.

\subsection{Additively Coupled Sum Constrained Games}

In this subsection, we present the definition of ACSCG and
subsequently, we present several exemplary multi-user scenarios
which appertain to this new class of game.

\begin{definition}
A multi-user interaction
$\Gamma=\langle\mathcal{N},\mathcal{A},u\rangle$ is a ACSCG if it
satisfies the following assumptions:

\textbf{A1:} $\forall n\in \mathcal{N}$, action set
$\mathcal{A}_n\subseteq\mathcal{R}^K$ is defined to be\footnote{We
consider a sum constraint throughout the paper rather than a
weighted-sum constraint, because a weighted-sum constraint can be
easily converted to a sum constraint by rescaling $\mathcal{A}_n$.
Besides, we nontrivially assume that $\sum_{k=1}^Ka_{n,k}^{\max}\geq
M_n$.}
\begin{equation}
\mathcal{A}_n=\Big\{(a_n^1,a_n^2,\cdots,a_n^K)\ \big|\ a_n^k \in
[a_{n,k}^{\min},a_{n,k}^{\max}]\ \textrm{and} \
\sum_{k=1}^Ka_n^k\leq M_n \Big\}.
\end{equation}

\textbf{A2:} There exist $h_n^k:\mathcal{R}\rightarrow \mathcal{R}$,
$f_n^k:\mathcal{A}_{-n}\rightarrow \mathcal{R}$, and
$g_n^k:\mathcal{A}_{-n}\rightarrow \mathcal{R}$, $k=1,\ldots,K$,
such that
\begin{equation}
\label{eq:eqn2}
u_n(\mathbf{a})=\sum_{k=1}^K\Big[h_n^k\big(a_{n}^k+f_n^k(\mathbf{a}_{-n})\big)-g_n^k(\mathbf{a}_{-n})\Big],
\end{equation}
for all $\mathbf{a}\in\mathcal{A}$ and $n\in\mathcal{N}$.
$h_n^k(\cdot)$ is an increasing, twice differentiable, and strictly
concave function and $f_n^k(\cdot)$ and $g_n^k(\cdot)$ are both
twice differentiable.
\end{definition}

%, if maximizing, or a continuous decreasing and strictly convex
%function and $F^k_{nn}\leq0$, if minimizing.\footnote[3]{Without
%loss of generality, we assume that users maximize their utilities,
%$h_n(\cdot)$ is continuous increasing and strictly concave, and
%$F^k_{nn}\geq0$ unless specified.}

The ACSCG model defined by assumptions A1 and A2 covers a broad
class of multi-user interactions. Assumption A1 indicates that each
player's action set is a $K$-dimensional vector set and its action
vector is sum-constrained. This represents the communication
scenarios in which each user needs to determine its
multi-dimensional action in various channels or networks while the
total amount of resources it can consume is constrained. Assumption
A2 implies that each user's utility is separable and can be
represented by the summation of concave functions $h_n^k$ minus
``penalty'' functions $g_n^k$ across the $K$ dimensions. In
particular, within each dimension, the input of $h_n^k$ is an
additive combination of user $n$'s action $a_{n}^k$ and function
$f_n^k(\mathbf{a}_{-n})$ that depends on the remaining users' joint
action $\mathbf{a}_{-n}$. Since $a_{n}^k$ only appears in the
concave function $h_n^k$, it implies that each user's utility is
concave in its own action, i.e. diminishing returns per unit of user
$n$'s invested action $\mathbf{a}_n$, which is common for many
application scenarios in communication networks.

Summarizing, the key features of the game model defined by A1 and A2
include: each user's action is subject to a \textit{sum constraint};
users' utilities are impacted by \emph{additive combinations} of
$a_{n}^k$ and $f_n^k(\mathbf{a}_{-n})$ through concave functions
$h_n^k$. Therefore, we term the game $\Gamma$ that satisfies
assumptions A1 and A2 as ACSCG. In the following section, we present
several illustrative multi-user interaction examples that belong to
ACSCG.

\subsection{Examples of ACSCG}
\label{sc:ex} We present four examples that satisfy assumptions A1
and A2 and belong to ACSCG. The details of functions $h_n^k(\cdot)$,
$f_n^k(\cdot)$ and $g_n^k(\cdot)$ in each example are summarized in
Table \ref{tb:tb1}. For each example, Table \ref{tb:tb1} also
summarizes the applicable convergence conditions that will be
provided in the remaining parts of the paper.

\begin{example}\label{ex:ex1}
We first consider a simple two-user game with two-dimension action
spaces, i.e. $N=K=2$. The utility functions are given by\footnote{In
this example, since there are only two users, the subindex $-n$
denotes the user but $n$.}
\begin{displaymath}
%u_n(\mathbf{a})=\sqrt{a_n^1+\frac{(a_{-n}^1)^2}{4}-\frac{(a_{-n}^2)^2}{9}}+
%\sqrt{a_n^2-\frac{(a_{-n}^1)^2}{5}+\frac{(a_{-n}^2)^2}{2}}
u_n(\mathbf{a})=-\exp\Bigl\{-a_n^1-\sqrt{(a_{-n}^1)^2+1}+\sqrt{(a_{-n}^2)^2+1}\Bigr\}
-\exp\Bigl\{-a_n^2+\sqrt{(a_{-n}^1)^2+1}-\sqrt{(a_{-n}^2)^2+1}\Bigr\},
\end{displaymath}
for $n=1,2$. The resource constraints are $\sum_{k=1}^2a_n^k\leq
M_n$ in which $M_n>0$ and $a_n^k\geq 0$ for $\forall n,k$.
\end{example}

\begin{example}\label{ex:ex2} \emph{(Power control in frequency-selective
Gaussian interference channel \cite{Yu_JSAC}\cite{Scutari})} There
are $N$ transmitter and receiver pairs in the system. The entire
frequency band is divided into $K$ frequency bins. In frequency bin
$k$, the channel gain from transmitter $i$ to receiver $j$ is
denoted as $H_{ij}^k$, where $k=1,2,\cdots,K$. Similarly, denote the
noise power spectral density (PSD) that receiver $n$ experiences as
$\sigma_n^k$ and player $n$'s transmit PSD as $P_n^k$. The action of
user $n$ is to select its transmit power $\mathbf{P}_n=[P_n^1\ P_n^2
\cdots P_n^K]$ subject to its power constraint: $ \sum_{k=1}^K
{P_n^k}\leq \textrm{P}_n^{\max}$. For a fixed $\mathbf{P}_n$, if
treating its interference as noise, user $n$ can achieve the
following data rate:
\begin{equation} \label{eq:eqn3}
\begin{gathered}
r_n(\mathbf{P})=\sum_{k=1}^K\log_2\Big(1+\frac{H_{nn}^kP_n^k}{\sigma_n^k+\sum_{m\neq
n} H_{mn}^kP_m^k}\Big) \qquad\qquad\qquad\qquad\qquad\qquad \\
=\sum_{k=1}^K\Big(\log_2(\sigma_n^k+\sum_{m=1}^N
H_{mn}^kP_m^k)-\log_2(\sigma_n^k+\sum_{m\neq n} H_{mn}^kP_m^k)\Big).
\end{gathered}
\end{equation}
\end{example}

\begin{table}
\centering \caption{Examples \ref{ex:ex1}-\ref{ex:ex4} as ACSCG.}
\begin{tabular}{|c||c|c|c|c|} \hline
\multirow{2}{*}{Examples} & \multirow{2}{*}{$f_n^k(\mathbf{a}_{-n})$} & \multirow{2}{*}{$h_n^k(x)$} & \multirow{2}{*}{$g_n^k(\mathbf{a}_{-n})$} & Convergence\\
& & & & conditions\\ \hline \hline
\multirow{2}{*}{Example \ref{ex:ex1}} & $f_n^1(\mathbf{a}_{-n})=\sqrt{(a_{-n}^1)^2+1}-\sqrt{(a_{-n}^2)^2+1}$ & \multirow{2}{*}{$-e^{-x}$} & \multirow{2}{*}{0} & \multirow{2}{*}{(\ref{C:C4})} \\
& $f_n^2(\mathbf{a}_{-n})=\sqrt{(a_{-n}^2)^2+1}-\sqrt{(a_{-n}^1)^2+1}$ & & &\\
\hline
%Example \ref{ex:ex1} & $f_n^1(\mathbf{a}_{-n})=\frac{(a_{-n}^1)^2}{4}-\frac{(a_{-n}^2)^2}{9}$,$f_n^2(\mathbf{a}_{-n})=\frac{(a_{-n}^2)^2}{2}-\frac{(a_{-n}^1)^2}{5}$ & $\sqrt{x}$ & 0 \\
%\hline
\multirow{2}{*}{Example \ref{ex:ex2}} & \multirow{2}{*}{$\sum\limits_{m\neq n} \frac{H_{mn}^k}{H_{nn}^k}P_m^k$} & \multirow{2}{*}{$\log_2(\sigma_n^k+H_{nn}^kx)$} & \multirow{2}{*}{$\log_2(\sigma_n^k+\sum\limits_{m\neq n} H_{mn}^kP_m^k)$} & \multirow{2}{*}{(\ref{C:C1})-(\ref{C:C8})} \\
& & & & \\
%Example \ref{ex:ex2} & $\sum_{m\neq n} \frac{H_{mn}^k}{H_{nn}^k}P_m^k$ & $\log_2(\sigma_n^k+H_{nn}^kx)$ & $\log_2(\sigma_n^k+\sum_{m\neq n} H_{mn}^kP_m^k)$\\
\hline \multirow{2}{*}{Example \ref{ex:ex3}} &
\multirow{2}{*}{$\sum\limits_{m\neq n}
\frac{\upsilon_{mn}^k}{\upsilon_{nn}^k}\psi_m^k$} &
\multirow{2}{*}{$-\frac{1}{\mu_n^k-\upsilon_{nn}^kx}$} & \multirow{2}{*}{0} & \multirow{2}{*}{(\ref{C:C1})-(\ref{C:C8})} \\
& & & &\\
\hline
\multirow{2}{*}{Example \ref{ex:ex4}} & \multirow{2}{*}{$\sum\limits_{m\neq n} (\sum\limits_{j=1}^K \frac{\gamma(k-j)H_{mn}^k}{H_{nn}^k}P_m^k)$} & \multirow{2}{*}{$\log_2(\sigma_n^k+H_{nn}^kx)$} & \multirow{2}{*}{$\log_2\bigl(\sigma_n^k+H_{nn}^kf_n^k(\mathbf{a}_{-n})\bigr)$}& \multirow{2}{*}{(\ref{C:C4})-(\ref{C:C8})} \\
& & & &\\
\hline\end{tabular} \label{tb:tb1}
\end{table}

\begin{example}\label{ex:ex3} (\emph{Delay minimization in Jackson Networks \cite{Jackson}})
As an additional example, we consider a network of $N$ nodes. A
Poisson stream of external packets arrive at node $n$ with rate
$\psi_n$ and the input stream is split into $K$ traffic classes,
which are individually served by exponential servers. Denote node
$n$'s input rate and service rate for class $k$ as $\psi_n^k$ and
$\mu_n^k$ respectively. Therefore, the action of node $n$ is to
determine the rates for different traffic classes $\Psi_n=[\psi_n^1\
\psi_n^2 \cdots \psi_n^K]$ and the total rate is subject to the
minimum rate constraint: $\sum_{k=1}^K {\psi_n^k}\geq
\psi_n^{\min}$. The packets of the same traffic class constitute a
Jackson network in which Markovian routing is adopted: packets of
class $k$ completing service at node $m$ are routed to node $n$ with
probability $r_{mn}^k$ or exit the network with probability
$r_{m0}^k=1-\sum_{n=1}^Nr^k_{mn}$. Denote the arrival rate for class
$k$ at node $n$ as $\eta_n^k$. By Jackson's Theorem, we have
$\eta_n^k=\psi_n^k+\sum_{m=1}^N\eta_m^kr^k_{mn}, n=1,2,\cdots,K$.
Denote $[\textrm{R}^k]_{mn}=r^k_{nm}$,
$\Upsilon^k=(\textrm{I}-\textrm{R}^k)^{-1}$, and
$\upsilon^k_{mn}=[\Upsilon^k]_{nm}$. Equivalently, we have
$\eta_n^k=\sum_{m=1}^N\upsilon_{mn}^k\psi_m^k$. Each node aims to
minimize its total M/M/1 queueing delay incurred by accommodating
its traffic:
\begin{equation}
\label{eq:eqn4}
d_n(\mathbf{\Psi})=\sum_{k=1}^K\frac{1}{\mu_n^k-\sum_{m=1}^N\upsilon_{mn}^k\psi_m^k}.
\end{equation}
\end{example}

Example \ref{ex:ex3} can be shown to be a special case of ACSCG by
slightly transforming the action sets and utilities. We can define
user $n$'s action as $-\Psi_n$. For user $n$, the sum constraint
becomes $\sum_{k=1}^K -{\psi_n^k}\leq -\psi_n^{\min}$ and minimizing
$d_n(\mathbf{\Psi})$ is equivalent to maximizing
$-d_n(\mathbf{\Psi})$.

\begin{example}\label{ex:ex4}
\emph{(Asynchronous transmission in digital subscriber lines network
\cite{ASB})} The basic setting of this example is similar to that of
Example \ref{ex:ex2} except that inter-carrier interference (ICI)
exist among different frequency bins. Due to the loss of the
orthogonality, the interference that user $n$ experiences in
frequency bin $k$ is
\begin{equation}
f_n^k(\mathbf{P}_{-n})=\sum_{m\neq n}\Bigl(\sum_{j=1}^K
\gamma(k-j)H_{mn}^jP_m^j\Bigr),
\end{equation}
in which $\gamma(j)$ is the ICI coefficient that represents the
relative interference transmitted signal in a particular frequency
bin generates to its $j$th neighbor bin. In particular, it takes the
form
\begin{equation}
\gamma(j)=\left\{ \begin{array}{cc}
1,& \text{if $j=0$}\\
\frac{2}{K^2\sin^2(\frac{\pi}{K}j)},& -\frac{K}{2}\leq j\leq
\frac{K}{2}\ j\neq0.
\end{array} \right.
\end{equation}
It satisfies the symmetric and circular properties, i.e.
$\gamma(-j)=\gamma(j)=\gamma(K-j)$. User $n$'s achievable rate in
the presence of ICI is given by
\begin{equation}
r_n(\mathbf{P})=\sum_{k=1}^K\log_2\Biggl[1+\frac{H_{nn}^kP_{n}^k}{\sigma_n^k+\sum_{m\neq
n}\Bigl(\sum_{j=1}^K \gamma(k-j)H_{mn}^jP_m^j\Bigr)}\Biggr].
\end{equation}
\end{example}

%\begin{remark}\label{rm:rm1}
\subsection{Issues related to ACSCG}
%\textit{(Issues related to ACSCG)}
Since the ACSCG model represents a good abstraction of numerous
multi-user resource allocation problems, we aim to investigate the
convergence properties of various distributed algorithms in ACSCG
without and with real-time message passing.

ACSCG is a concave game\cite{Game_book}\cite{Rosen} and therefore,
it admits at least one pure NE. In practice, we want to provide the
sufficient conditions under which best response dynamics provably
and globally converges to a pure NE. However, the existing
literature, e.g. the diagonal strict concavity (DSC) conditions in
\cite{Rosen} and the supermodular game theory
\cite{Topkis_book}-\cite{Sgame_2}, does not provide such convergence
conditions for the general ACSCG model. For example, the DSC
conditions developed for general concave games do not guarantee the
convergence of best response dynamics \cite{Rosen}. Even if the
utility functions in ACSCG possess the supermodular type structure,
due to the sum constraint, the action set of each user is generally
not a sublattice\footnote{In supermodular games, for each player,
the action set is a nonempty and compact sublattice of
$\mathcal{R}^K$. We can verify that with the sum constraint,
$\mathcal{A}_n$ is usually not a sublattice of $\mathcal{R}^K$ by
taking the component-wise maximum.} of $\mathcal{R}^K$. Therefore,
the convergence results based on supermodular games cannot be
directly applied in ACSCG. On the other hand, if we want to maximize
the sum utility by enabling real-time message passing among users,
we also note that, the utility $u_n$ is not necessarily jointly
concave in $\textbf{a}$ because of the existence of $g_n^k(\cdot)$.
Therefore, the existing algorithms developed for the convex NUM are
not immediately applicable either.

In fact, a unique feature of the ACSCG is that different users'
actions are \emph{additively coupled} in $h_n^k(\cdot)$ and each
user's action space is \emph{sum-constrained}. In the following
sections, we will fully explore these specific structures and
address the convergence properties of various distributed algorithms
in two different scenarios. Specifically, Section \ref{sc:non-cp}
investigates the scenarios in which each user $n$ can only observe
$\{f_n^k(\mathbf{a}_{-n})\}_{k=1}^K$ and cannot exchange any
information with any other user. Section \ref{sc:cp} focuses on the
scenarios in which each user $n$ is able to announce and receive
information in real-time to and from the remaining users about
$\frac{\partial u_n(\mathbf{a})}{\partial a_m^k}$ and
$\frac{\partial u_m(\mathbf{a})}{\partial a_n^k}$, $\forall m\neq n,
k=1,\ldots,K$.
%For example, we know that
%\begin{equation}
%\label{eq:eqn5} \frac{\partial^2 u_n}{\partial a_n^k\partial
%a_m^k}=F_{nn}^kF_{mn}^kh^{''k}_n\big((\mathbf{F}_n^k)^T\mathbf{a}^k+\alpha_n^k\big).
%\end{equation}
%Since $h_n^k(\cdot)$ is strictly concave and $F_{nn}^k\geq0$, the
%sign of $\frac{\partial^2 u_n}{\partial a_n^k\partial a_m^k}$
%depends on $F_{mn}^k$. Therefore, if $\exists m,n,k,m',n',k'$ such
%that $F_{mn}^kF_{m'n'}^{k'}\leq0$, the utilities in ACSCG do not
%necessarily have the supermodular type structure (or the more
%general S-modular type structure) \cite{Topkis_book}-\cite{Sgame_2}.
%Moreover, even if $F_{mn}^kF_{m'n'}^{k'}\geq0, \forall m\neq
%n,m'\neq n',k,k'$,
%\end{remark}

\section{Scenario I: no message exchange among users}
\label{sc:non-cp} In communication scenarios where users cannot
exchange messages to achieve coordination, the participating users
can simply choose actions to selfishly maximize their individual
utility functions $u_n(\mathbf{a})$ without taking into account the
utility degradation caused to the other users. In particular, each
user individually solves the following optimization program:
\begin{equation}\label{eq:eqn6}
\max_{\textbf{a}_n\in \mathcal{A}_n}u_n(\mathbf{a}).
%\textrm{s.t.} \ \ \sum_{k=1}^Ka_n^k\leq M_n.
\end{equation}
The steady state outcome of such a multi-user interaction is usually
characterized as a NE, at which given the other users' actions, no
user can increase its utility alone by unilaterally changing its
action. It is worth pointing out that, since there is no
coordination signal among users, NE generally does not lead to a
Pareto-optimal solution. Section IV will discuss distributed
algorithms in which users exchange coordination signals in order to
improve the system efficiency.

\subsection{Properties of Best Response Dynamics in ACSCG}
\label{sc:l}To better understand the key properties of the ACSCG, in
this subsection, we first focus on the scenarios in which
$f_n^k(\mathbf{a}_{-n})$ is the linear combination of the remaining
users' action in the same dimension $k$, i.e.
\begin{equation}\label{eq:simplef}
f_n^k(\mathbf{a}_{-n})=\sum_{m\neq n}F_{mn}^ka_m^k
\end{equation}
and $F_{mn}^k\in \mathcal{R}$, $\forall m,n,k$. Specifically, both
Example \ref{ex:ex2} and \ref{ex:ex3} in Table \ref{tb:tb1} belong
to this category. In Section \ref{sc:nl}, we will extend the results
derived for the functions $f_n^k(\mathbf{a}_{-n})$ defined in
(\ref{eq:simplef}) to general $f_n^k(\mathbf{a}_{-n})$.

Since $h_n^k(\cdot)$ is concave, the objective in (\ref{eq:eqn6}) is
a concave function in $a_n^k$ when the other users' actions
$\textbf{a}_{-n}$ are fixed. To find the globally optimal solution
of the problem in (\ref{eq:eqn6}), we can first form its Lagrangian
\begin{equation}
\label{eq:eqn7} L_n(\mathbf{a}_n,
\lambda)=u_n(\mathbf{a})+\lambda(M_n-\sum_{k=1}^Ka_n^k),
\end{equation}
in which $a_n^k \in[a_{n,k}^{\min},a_{n,k}^{\max}]$. By taking the
first derivatives of (\ref{eq:eqn7}), we have
\begin{equation}
\label{eq:eqn8} \frac{\partial L_n(\mathbf{a}_n, \lambda)}{\partial
a_n^k}=\frac{\partial h_n^{k}(a_n^k+\sum_{m\neq
n}F_{mn}^ka_m^k)}{\partial a_n^k} -\lambda=0.
\end{equation}
Denote
\begin{equation}
\label{eq:eqn9} l_n^k(\mathbf{a}_{-n},\lambda)\triangleq\Big[
\Bigl\{\frac{\partial h_n^{k}}{\partial
x}\Bigr\}^{-1}(\lambda)-\sum_{m\neq
n}F_{mn}^ka_m^k\Big]^{a_{n,k}^{\max}}_{a_{n,k}^{\min}},
\end{equation}
in which $\bigl\{\frac{\partial h_n^{k}}{\partial x}\bigr\}^{-1}$ is
the inverse function\footnote{If $\nexists \ x=x^*$ such that
$\frac{\partial h_n^{k}}{\partial x}|_{x=x^*}=\lambda$, we let
$\bigl\{\frac{\partial h_n^{k}}{\partial
x}\bigr\}^{-1}(\lambda)=-\infty$.} of $\frac{\partial
h_n^{k}}{\partial x}$ and $[x]^a_b=\max\{\min\{x,a\},b\}$. The
optimal solution of (\ref{eq:eqn6}) is given by
$a_n^{*k}=l_n^k(\mathbf{a}_{-n},\lambda^*)$, where the Lagrange
multiplier $\lambda^*$ is chosen to satisfy the sum constraint
$\sum_{k=1}^K a_n^{*k}=M_n$.

We define the best response operator $B_n^k(\cdot)$ as
\begin{equation}
\label{eq:eqn10}
B_n^k(\mathbf{a}_{-n})=l_n^k(\mathbf{a}_{-n},\lambda^*).
\end{equation}

We consider the dynamic adjustment process in which users revise
their actions over time based on their observations about their
opponents. A well-known candidate for such adjustment processes is
the so-called best response dynamics. In the best response
algorithm, each user updates its action using the best response
strategy that maximizes its utility function in (\ref{eq:eqn2}). We
consider two types of update orders, including sequential update and
parallel update. Specifically, in sequential update, individual
players iteratively optimize in a circular fashion with respect to
their own actions while keeping the actions of their opponents
fixed. Formally, at stage $t$, user $n$ chooses its action according
to
\begin{equation}
\label{eq:eqn11}
a_n^{k,t}=B_n^k([\mathbf{a}^{t}_{1},\ldots,\mathbf{a}^{t}_{n-1},\mathbf{a}^{t-1}_{n+1},\ldots,\mathbf{a}^{t-1}_{N}]).
\end{equation}
On the other hand, players adopting the parallel update revise their
actions at stage $t$ according to
\begin{equation}
\label{eq:eqn12} a_n^{k,t}=B_n^k(\mathbf{a}^{t-1}_{-n}).
\end{equation}

We obtain several sufficient conditions under which best response
dynamics converges. Similar convergence conditions are proved in
\cite{Chung}-\cite{Scutari} for Example \ref{ex:ex2} in which
$h_n^k(x)=\log_2(\sigma_n^k+H_{nn}^kx)$. We consider more general
functions $h_n^k(\cdot)$ and further extend the convergence
conditions in \cite{Chung}-\cite{Scutari}. The key differences among
all the sufficient conditions which will be provided in this section
are summarized in Table \ref{tb:tb2}.

\begin{table}
\centering \caption{Comparison among conditions
(\ref{C:C1})-(\ref{C:C6}).}
\begin{tabular}{|c||c|c|c|c|} \hline
Conditions & Assumptions about $f_n^k(\mathbf{a}_{-n})$ & $h_n^k(x)$ & Measure of residual error $\textbf{a}_n^{t+1}-\textbf{a}_n^{t}$ & Contraction factor\\
\hline \hline
(\ref{C:C1}) & (\ref{eq:simplef}) & A2 & 1-norm & $2\rho(\textbf{T}^{\max})$ \\
\hline\multirow{2}{*}{(\ref{C:C2})} & (\ref{eq:simplef}) and $F_{mn}^k$ have & \multirow{2}{*}{A2} & \multirow{2}{*}{1-norm} & \multirow{2}{*}{$\rho(\textbf{T}^{\max})$} \\
 & the same sign for $\forall k, m\neq n$ & & &\\
\hline(\ref{C:C3}) & (\ref{eq:simplef}) & (\ref{eq:eqn14}) & weighted Euclidean norm & $\rho(\textbf{S}^{\max})$ \\
\hline
(\ref{C:C4}) & general & A2 & 1-norm & $2\rho(\bar{\textbf{T}}^{\max})$ \\
\hline \multirow{2}{*}{(\ref{C:C5})} & $\frac{\partial
f_n^k(\textbf{a}_{-n})}{\partial a_m^{k'}}$ have the same sign & \multirow{2}{*}{A2} & \multirow{2}{*}{1-norm} & \multirow{2}{*}{$\rho(\bar{\textbf{T}}^{\max})$} \\
& for $\forall \textbf{a}\in\mathcal{A},k,k',m\neq n$ & & &\\
\hline(\ref{C:C6}) & general & (\ref{eq:eqn14}) & weighted Euclidean norm & $\rho(\bar{\textbf{S}}^{\max})$ \\
\hline\end{tabular} \label{tb:tb2}
\end{table}

\subsubsection{General $h_n^k(\cdot)$}
The first sufficient condition is developed for the general cases in
which the functions $h_n^k(\cdot)$ in the utilities $u_n(\cdot)$ are
specified in assumption A2. Define
\begin{equation}
\label{eq:Tmax}
[\textbf{T}^{\max}]_{mn}\triangleq\left\{ \begin{array}{cl}
\max_{k}|F_{mn}^k|,& \text{if $m\neq n$}\\
0,& \text{otherwise}.
\end{array} \right.
\end{equation} and let $\rho(\textbf{T}^{\max})$ denote the spectral
radius of the matrix $\textbf{T}^{\max}$.

\begin{theorem}\label{th:th1}
If
\begin{equation}
\label{C:C1} \tag{C1} \rho(\textbf{T}^{\max})<\frac{1}{2},
\end{equation}
then there exists a unique NE in game $\Gamma$ and best response
dynamics converges linearly to the NE, for any set of initial
conditions belonging to $\mathcal{A}$ with either sequential or
parallel updates.
\end{theorem}

\textit{Proof}: This theorem is proved by showing that the best
response dynamics defined in (\ref{eq:eqn11}) and (\ref{eq:eqn12})
is a contraction mapping under (\ref{C:C1}). See Appendix A for
details. $\blacksquare$

In multi-user communication applications, it is common to have games
of \textit{strategic complements} (or \textit{strategic
substitutes}), i.e. the marginal returns to any one component of the
player's action rise with increases (or decreases) in the components
of the competitors' actions \cite{complements}. For instance, in
Examples \ref{ex:ex2} and \ref{ex:ex4}, increasing user $n$'s
transmitted power creates stronger interference to the other users
and decreases their marginal achievable rates. Similarly, in Example
\ref{ex:ex3}, increasing node $n$'s input traffic rate congests all
the servers in the network and increases the marginal queueing
delay. Mathematically, if $u_n$ is twice differentiable, strategic
complementarities (or strategic substitutes) can be described as
\begin{equation}
\frac{\partial^2 u_n(\mathbf{a}_n,\mathbf{a}_{-n})}{\partial
a_n^j\partial a_m^k}\geq0,\ \forall m\neq n,j,k, (\textrm{or} \
\frac{\partial^2 u_n(\mathbf{a}_n,\mathbf{a}_{-n})}{\partial
a_n^j\partial a_m^k}\leq0,\ \forall m\neq n,j,k).
\end{equation}
We can verify that Examples \ref{ex:ex2}, \ref{ex:ex3}, and
\ref{ex:ex4} are games with strategic substitutes. For the ACSCG
models that exhibit strategic complementarities (or strategic
substitutes), the following theorem further relaxes condition
(\ref{C:C1}).

\begin{theorem}\label{co:co1} Let $\Gamma$ be an ACSCG with strategic complementarities
(or strategic substitutes), i.e. $F_{mn}^k\leq0$, $\forall k,m\neq
n$, (or $F_{mn}^k\geq0$, $\forall k,m\neq n$). If
\begin{equation}
\label{C:C2} \tag{C2} \rho(\textbf{T}^{\max})<1,
\end{equation}
then there exists a unique NE in game $\Gamma$ and best response
dynamics converges linearly to the NE, for any set of initial
conditions belonging to $\mathcal{A}$ with either sequential or
parallel updates.
\end{theorem}

\textit{Proof}: This theorem is proved by adapting the proof of
Theorem \ref{th:th1}. See Appendix B. $\blacksquare$

\begin{remark}\label{rm:rm2}
\textit{(Implications of conditions (\ref{C:C1}) and (\ref{C:C2}))}
Theorem \ref{th:th1} and Theorem \ref{co:co1} give sufficient
conditions for best response dynamics to globally converge to a
unique fixed point. Specifically, $\max_{k}|F_{mn}^k|$ can be
regarded as a measure of the strength of the mutual coupling between
user $m$ and $n$. The intuition behind (\ref{C:C1}) and (\ref{C:C2})
is that, the weaker the coupling among different users is, the more
likely that best response dynamics converges. Consider the extreme
case in which $F_{mn}^k=0, \forall k,m\neq n$. Since each user's
best response is not impacted by the remaining users' action
$\textbf{a}_{-n}$, the convergence is immediately achieved after a
single best-response iteration. If no restriction is imposed on
$F_{mn}^k$, Theorem \ref{th:th1} specifies a mutual coupling
threshold under which best response dynamics provably converge. The
proof of Theorem \ref{th:th1} can be intuitively interpreted as
follows. We regard every best response update as the users' joint
attempt to approach the NE. Due to the linear coupling structure in
(\ref{eq:simplef}), user $n$'s best response in (\ref{eq:eqn9})
contains a term $\sum_{m\neq n}F_{mn}^ka_m^k$ that is a linear
combination of $\textbf{a}_{-n}$. As a result, the residual error
$\big|\textbf{a}_n^{t+1}-\textbf{a}_n^t\big|_1$, which is the 1-norm
distance between the updated action profile $\textbf{a}_n^{t+1}$ and
the current action profile $\textbf{a}_n^t$, can be upper-bounded
using linear combinations of
$\big|\textbf{a}_m^t-\textbf{a}_m^{t-1}\big|_1$ in which $m\neq n$.
Recall that $F_{mn}^k$ can be either positive or negative. We also
note that, if $\textbf{a}_m^{t}\neq\textbf{a}_m^{t-1}$,
$\textbf{a}_m^{t}-\textbf{a}_m^{t-1}$ contains both positive and
negative terms due to the sum-constraint. In the worst case, the
distance $\big|\textbf{a}_n^{t+1}-\textbf{a}_n^{t}\big|_1$ is
maximized if $\big\{F_{mn}^k\big\}$ and
$\big\{a_m^{k,t}-a_m^{k,t-1}\big\}$ are co-phase multiplied and
additively summed, i.e.
$F_{mn}^k\big(a_m^{k,t}-a_m^{k,t-1}\big)\geq0$, for $\forall
k=1,\ldots,K, m\neq n$. After an iteration, all users except $n$
contributes to user $n$'s residual error at stage $t+1$ up to
$\sum_{m\neq
n}2\max_k\big|F_{mn}^k\big|\big|\textbf{a}_m^{t}-\textbf{a}_m^{t-1}\big|_1$.
Under condition (\ref{C:C1}), it is guaranteed that the residual
error contracts with respect to the special norm defined in
(\ref{eq:specialnorm}). Theorem \ref{co:co1} focuses on the
situations in which the signs of $F_{mn}^k$ are the same, $\forall
m\neq n,k$. In this case, $\big\{F_{mn}^k\big\}$ and
$\big\{a_m^{k,t}-a_m^{k,t-1}\big\}$ cannot be co-phase multiplied.
Therefore, the region of convergence enlarges and hence, condition
(\ref{C:C2}) stated in Theorem \ref{co:co1} is weaker than condition
(\ref{C:C1}) in Theorem \ref{th:th1}.
\end{remark}

\begin{remark}
\textit{(Relation to the results in references
\cite{Chung}-\cite{Scutari})} Similar to \cite{Chung}\cite{ASB}, our
proofs choose 1-norm as the distance measure for the residual errors
$\textbf{a}_n^{t+1}-\textbf{a}_n^{t}$ after each best-response
iteration. However, by manipulating the inequalities in a different
way, condition (\ref{C:C2}) is more general than the results in
\cite{Chung}\cite{ASB}, where they require
$\max_kF_{mn}^k<\frac{1}{N-1}$. Interestingly, condition
(\ref{C:C2}) recovers the result obtained in \cite{Scutari} where it
is proved by choosing the Euclidean norm as the distance measure for
the residual errors $\textbf{a}_n^{t+1}-\textbf{a}_n^{t}$ after each
best-response iteration. However, the approach in \cite{Scutari}
using the Euclidean norm only applies to the scenarios in which
$h_n^k(\cdot)$ is a logarithmic function. We prove that condition
(\ref{C:C2}) applies to any $h_n^k(\cdot)$ that is increasing and
strictly concave.
\end{remark}

\subsubsection{A special class of $h_n^k(\cdot)$}
In addition to conditions (\ref{C:C1}) and (\ref{C:C2}), we also
develop a sufficient convergence condition for a family of utility
functions parameterized by a negative number $\theta$. In
particular, $h_n^k(\cdot)$ satisfies\footnote{If
$\alpha_n^{k}+F_{nn}^kx\leq0$, we let $h_n^k(x)=-\infty$. We assume
for this class of $h_n^k(\cdot)$ that for $\forall
\mathbf{a}_{-n}\in\mathcal{A}_{-n}$, there exists
$\mathbf{a}_{n}\in\mathcal{A}_{n}$ such that
$\alpha_n^{k}+F_{nn}^kx>0$ for $\forall n,k$.}
%\begin{equation}
%\label{eq:eqn13} \bigl[h_n^{k}\bigr]'(x)=(\alpha_n^{k}+F_{nn}^kx)^{\theta}, \theta<0.
%\end{equation}
%The interpretation of this type of utilities has been addressed in
%\cite{ToNUtilities}. It is shown that varying the parameter $\theta$
%leads to different types of fairness across
%$(\mathbf{F}_n^k)^T\mathbf{a}^k+\alpha_n^k$ for all $k$. In
%particular, $\theta=-1$ corresponds to the proportional fairness; if
%$\theta=-2$, then harmonic mean fairness; and if $\theta=-\infty$,
%then max-min fairness. Given the functional expression of the first
%derivative $\bigl[h_n^{k}\bigr]'(\cdot)$, the original function $h_n^k(\cdot)$
%takes the form
\begin{equation} \label{eq:eqn14}
h_n^k(x)=\left\{ \begin{array}{cl}
\log(\alpha_n^{k}+F_{nn}^kx),& \text{if $\theta=-1$},\\
\frac{(\alpha_n^{k}+F_{nn}^kx)^{\theta+1}}{\theta+1},& \text{if
$-1<\theta<0$ or $\theta<-1$}.
\end{array} \right.
\end{equation}
and $\alpha_n^{k}\in\mathcal{R}$ and $F_{nn}^{k}>0$. The
interpretation of this type of utilities has been addressed in
\cite{ToNUtilities}. It is shown that varying the parameter $\theta$
leads to different types of fairness across
$\alpha_n^{k}+F_{nn}^k(a_{n}^k+\sum_{m\neq n}F_{mn}^ka_m^k)$ for all
$k$. In particular, $\theta=-1$ corresponds to the proportional
fairness; if $\theta=-2$, then harmonic mean fairness; and if
$\theta=-\infty$, then max-min fairness. We can see that, Examples
\ref{ex:ex2} and \ref{ex:ex3} are special cases of this type of
utility functions. In these cases, best response dynamics in
equation (\ref{eq:eqn9}) is reduced to
\begin{equation} \label{eq:eqn15}
l_n^k(\mathbf{a}_{-n},\lambda)=\Big[\big(\frac{1}{F_{nn}^k}\big)^{1+\frac{1}{\theta}}\lambda^{\frac{1}{\theta}}
-\frac{\alpha_n^k}{F_{nn}^k}-\sum_{m\neq n}F_{mn}^ka_m^k\Big]^{a_{n,k}^{\max}}_{a_{n,k}^{\min}},
\end{equation}
Define
\begin{equation} \label{eq:eqn16}
[\textbf{S}^{\max}]_{mn}\triangleq\left\{ \begin{array}{cl}
\frac{\sum_{k=1}^K(F_{mm}^k)^{1+\frac{1}{\theta}}}{\sum_{k=1}^K(F_{nn}^k)^{1+\frac{1}{\theta}}}\max_{k}\Bigl\{|F_{mn}^k|\Bigl(\frac{F_{nn}^k}{F_{mm}^k}\Bigr)^{1+\frac{1}{\theta}}\Bigr\},& \text{if $m\neq n$}\\
0,& \text{otherwise}.
\end{array} \right.
\end{equation} For the class of utility
functions in (\ref{eq:eqn14}), Theorem \ref{th:th2} gives a
sufficient condition that guarantees the convergence of the best
response dynamics defined in (\ref{eq:eqn15}).

\begin{theorem}\label{th:th2} For $h_n^{k}(\cdot)$ defined in (\ref{eq:eqn14}), if
\begin{equation}
\label{C:C3} \tag{C3} \rho(\textbf{S}^{\max})<1,
\end{equation}
then there exists a unique NE in game $\Gamma$ and best response
dynamics converges linearly to the NE, for any set of initial
conditions belonging to $\mathcal{A}$ and with either sequential or
parallel updates.
\end{theorem}

\textit{Proof}: It can be proved by showing that the best response
dynamics defined in (\ref{eq:eqn15}) is a contraction mapping with
respect to the weighted Euclidean norm. See Appendix C for details.
$\blacksquare$

\begin{remark}
\textit{(Relation between conditions (\ref{C:C3}) and the results in
reference \cite{Scutari})} For aforementioned Example \ref{ex:ex2},
Scutari et al. established in \cite{Scutari} a sufficient condition
under which the iterative water-filling algorithm converges. The
iterative water-filling algorithm essentially belongs to best
response dynamics. Specifically, in \cite{Scutari}, Shannon's
formula leads to $\theta=-1$ and cross channel coefficients satisfy
$F_{mn}^k\geq0, \forall k,m\neq n$. Equation (\ref{eq:eqn15})
reduces to the water-filling formula
\begin{equation} \label{eq:eqnwf}
l_n^k(\mathbf{a}_{-n},\lambda)=\Big[\frac{1}{\lambda}-\frac{\alpha_n^k}{F_{nn}^k}-
\sum_{m\neq n}F_{mn}^ka_m^k\Big]^{a_{n,k}^{\max}}_{a_{n,k}^{\min}},
\end{equation}
and $[\textbf{S}^{\max}]_{mn}=\max_{k}F_{mn}^k$. By choosing the
weighted Euclidean norm as the distance measure for the residual
errors $\textbf{a}_n^{t+1}-\textbf{a}_n^{t}$ after each
best-response iteration, Theorem \ref{th:th2} generalizes the
results in \cite{Scutari} for the family of utility functions
defined in (\ref{eq:eqn14}).
\end{remark}

\begin{figure}
\centering
\includegraphics[width=0.7\textwidth]{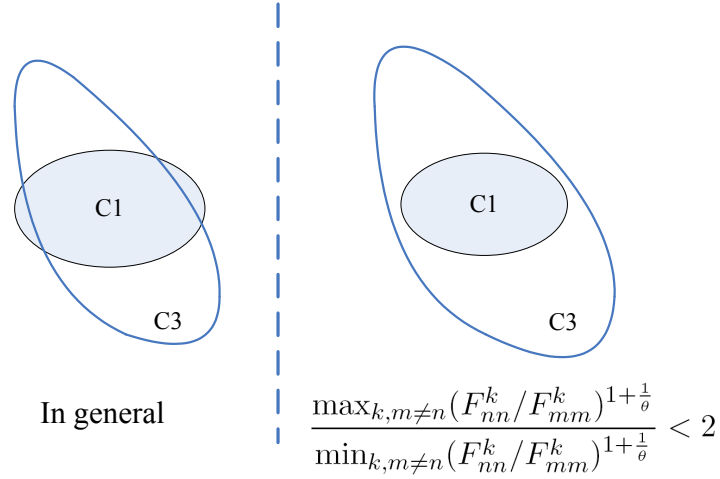}
\caption{Relation between (\ref{C:C1}) and (\ref{C:C3}).}
\label{fg:fg1}
\end{figure}

\begin{remark}\label{rm:rm3}
\textit{(Relation between conditions (\ref{C:C1}), (\ref{C:C2}) and
(\ref{C:C3}))} The connections and differences between conditions
(\ref{C:C1}), (\ref{C:C2}) and (\ref{C:C3}) are summarized in Table
\ref{tb:tb2}. We have addressed the implications of (\ref{C:C1}) and
(\ref{C:C2}) in Remark \ref{rm:rm2}. Now we discuss their relation
with (\ref{C:C3}). First of all, condition (\ref{C:C1}) is proposed
for general $h_n^k(\cdot)$ and condition (\ref{C:C3}) is proposed
for the class of utility functions defined in (\ref{eq:eqn14}).
However, Theorem \ref{th:th1} and Theorem \ref{th:th2} individually
establish the fact that best response dynamics is a contraction map
by selecting different vector and matrix norms. Therefore, in
general, (\ref{C:C1}) and (\ref{C:C3}) do not immediately imply each
other. Note that
$[\textbf{S}^{\max}]_{mn}\leq\zeta_{mn}\cdot\max_{k}|F_{mn}^k|$ in
which $\zeta_{mn}$ satifies
\begin{equation}\label{eq:zeta}
\zeta_{mn}=\frac{\sum_{k=1}^K(F_{mm}^k)^{1+\frac{1}{\theta}}}{\sum_{k=1}^K(F_{nn}^k)^{1+\frac{1}{\theta}}}\cdot
\max_{k}\frac{(F_{nn}^k)^{1+\frac{1}{\theta}}}{(F_{mm}^k)^{1+\frac{1}{\theta}}}\in\Big[1,\frac{\max_{k}(F_{nn}^k/F_{mm}^k)^{1+\frac{1}{\theta}}}{\min_{k}(F_{nn}^k/F_{mm}^k)^{1+\frac{1}{\theta}}}\Big].
\end{equation}
The physical interpretation of $\zeta_{mn}$ is the similarity
between the preferences of user $m$ and $n$ across the total $K$
dimensions of their action spaces. Recall that both
$\textbf{S}^{\max}$ and $\textbf{T}^{\max}$ are non-negative
matrices and $\textbf{S}^{\max}$ is element-wise less than or equal
to $\max_{m\neq n}\zeta_{mn}\textbf{T}^{\max}$. By the property of
non-negative matrix and condition (\ref{C:C1}), we can conclude
$\rho(\textbf{S}^{\max})\leq\rho(\max_{m\neq
n}\zeta_{mn}\textbf{T}^{\max})<\max_{m\neq n}\frac{\zeta_{mn}}{2}$.
The relation between (\ref{C:C1}) and (\ref{C:C3}) is pictorially
illustrated in Fig. \ref{fg:fg1}. Specifically, if users have
similar preference in their available actions and the upper bound of
$\zeta_{mn}$ that measures the difference of their preferences is
below the following threshold:
\begin{equation}
\label{eq:ratio} \frac{\max_{k,m\neq
n}(F_{nn}^k/F_{mm}^k)^{1+\frac{1}{\theta}}}{\min_{k,m\neq
n}(F_{nn}^k/F_{mm}^k)^{1+\frac{1}{\theta}}}<2,
\end{equation}
we know that (\ref{C:C1}) implies (\ref{C:C3}) in this situation
because
$\rho(\textbf{S}^{\max})<\max_{m,n}\zeta_{mn}\cdot\rho(\textbf{T}^{\max})<2\cdot\frac{1}{2}=1$.
We also would like to point out that, the LHS of (\ref{eq:ratio}) is
a function of $\theta$ and the LHS $\equiv1$ if $\theta=-1$. When
$\theta=-1$, $\textbf{T}^{\max}$ coincides with $\textbf{S}^{\max}$.
Mathematically, in this case, (\ref{C:C3}) is actually more general
than (\ref{C:C2}), because it still holds even if coefficients
$F_{mn}^k$ have different signs.
\end{remark}

\subsection{Extensions to General $f_n^k(\cdot)$}
\label{sc:nl} As a matter of fact, the results above can be extended
to the more general situations
%in which the mutual coupling among
%users are non-linear. Consider the following utility function for
%user $n$:
%\begin{equation}
%\label{eq:eqn17}
%u_n(\mathbf{a})=\sum_{k=1}^K\Big[h_n^k\big(F_{nn}^ka_n^k+\sum_{m\neq
%n}f_{mn}^k(a_m^k)+\alpha_n^k\big)+g_n^k(\mathbf{a}_{-n}^k)\Big],
%\end{equation}
in which $f_n^k(\cdot)$ is a nonlinear differentiable function,
$\forall n,k$ and its input $\mathbf{a}_{-n}$ consists of the
remaining users' action from all the dimensions. Accordingly,
equation (\ref{eq:eqn9}) becomes
\begin{equation}
\label{eq:newBR}l_n^k(\mathbf{a}_{-n},\lambda)\triangleq\Big[
\Bigl\{\frac{\partial h_n^{k}}{\partial
x}\Bigr\}^{-1}(\lambda)-f_n^k(\mathbf{a}_{-n})\Big]^{a_{n,k}^{\max}}_{a_{n,k}^{\min}}.
\end{equation} The conclusions in Theorem \ref{th:th1}, \ref{co:co1}, and
\ref{th:th2} can be further extended as Theorem \ref{th:th3}, and
\ref{co:co2}, \ref{th:th4} that are listed below. We only provide
the proof of Theorem \ref{th:th3} in Appendix D. The detailed proofs
of Theorem \ref{co:co2} and \ref{th:th4} are omitted because they
can be proven similarly as Theorem \ref{th:th3}.

For general $f_n^k(\cdot)$, we denote
\begin{equation}
\label{eq:Tbarmax} [\bar{\textbf{T}}^{\max}]_{mn}\triangleq\left\{
\begin{array}{cl}
\max_{\textbf{a}\in\mathcal{A},k'}\sum_{k=1}^K\Big|\frac{\partial
f_n^k(\textbf{a}_{-n})}{\partial a_m^{k'}}\Big|,& \text{if $m\neq n$}\\
0,& \text{otherwise}.
\end{array} \right.
\end{equation}
Besides, for $h_n^{k}(\cdot)$ defined in (\ref{eq:eqn14}), we define
\begin{equation}
\label{eq:Sbarmax} [\bar{\textbf{S}}^{\max}]_{mn}\triangleq\left\{
\begin{array}{cl}
\frac{\sum_{k=1}^K(F_{mm}^k)^{1+\frac{1}{\theta}}}{\sum_{k=1}^K(F_{nn}^k)^{1+\frac{1}{\theta}}}
\max_{\textbf{a}\in\mathcal{A},k'}\Bigl\{\sum_{k=1}^K\Big|\frac{\partial
f_n^k(\textbf{a}_{-n})}{\partial a_m^{k'}}\Big|\Bigl(\frac{F_{nn}^{k'}}{F_{mm}^{k'}}\Bigr)^{1+\frac{1}{\theta}}\Bigr\},& \text{if $m\neq n$}\\
0,& \text{otherwise}.
\end{array} \right.
\end{equation}

\begin{theorem}\label{th:th3}
If
\begin{equation}
\label{C:C4} \tag{C4} \rho(\bar{\textbf{T}}^{\max})<\frac{1}{2},
\end{equation}
then there exists a unique NE in game $\Gamma$ and best response
dynamics converges linearly to the NE, for any set of initial
conditions belonging to $\mathcal{A}$ with either sequential or
parallel updates.
\end{theorem}

\textit{Proof}: This theorem can be proved by combining the proof of
Theorem \ref{th:th1} and the mean value theorem for vector-valued
functions. See Appendix D for details. $\blacksquare$

Similarly as in Theorem \ref{co:co1}, for the general ACSCG models
that exhibit strategic complementarities (or strategic substitutes),
we can further relax condition (\ref{C:C4}).

\begin{theorem}\label{co:co2} For $\Gamma$ with strategic complementarities
(or strategic substitutes), i.e. $\frac{\partial
f_n^k(\textbf{a}_{-n})}{\partial a_m^{k'}}\geq0, \forall m\neq
n,k,k',\textbf{a}\in\mathcal{A}$, (or $\frac{\partial
f_n^k(\textbf{a}_{-n})}{\partial a_m^{k'}}\leq0, \forall m\neq
n,k,k',\textbf{a}\in\mathcal{A}$), if
\begin{equation}
\label{C:C5} \tag{C5} \rho(\bar{\textbf{T}}^{\max})<1,
\end{equation}
then there exists a unique NE in game $\Gamma$ and best response
dynamics converges linearly to the NE, for any set of initial
conditions belonging to $\mathcal{A}$ with either sequential or
parallel updates.
\end{theorem}

\begin{theorem}\label{th:th4}
For $h_n^{k}(\cdot)$ defined in (\ref{eq:eqn14}), if
\begin{equation}
\label{C:C6} \tag{C6} \rho(\bar{\textbf{S}}^{\max})<1,
\end{equation}
then there exists a unique NE in game $\Gamma$ and best response
dynamics converges linearly to the NE, for any set of initial
conditions belonging to $\mathcal{A}$ with either sequential or
parallel updates.
\end{theorem}

\begin{remark}\label{rm:rm4}
\textit{(Implications of conditions (\ref{C:C4}), (\ref{C:C5}), and
(\ref{C:C6}))} Based on the mean value theorem, we know that the
upper bound of the additive sum of first derivatives
$\sum_{k=1}^K\Big|\frac{\partial f_n^k(\textbf{a}_{-n})}{\partial
a_m^{k'}}\Big|$ governs the maximum impact that user $m$'s action
can make over user $n$'s utility. As a result, Theorem \ref{th:th3},
Theorem \ref{co:co2}, and Theorem \ref{th:th4} indicate that
$\sum_{k=1}^K\Big|\frac{\partial f_n^k(\textbf{a}_{-n})}{\partial
a_m^{k'}}\Big|$ can be used to develop similar sufficient conditions
for the global convergence of best response dynamics. Table
\ref{tb:tb2} summarizes the connections and differences among all
the aforementioned conditions from (\ref{C:C1}) to (\ref{C:C6}). We
can verify that, for the linear function $f_n^k(\cdot)$ that is
defined in (\ref{eq:simplef}) and studied in Section \ref{sc:l},
$\forall \textbf{a}\in \mathcal{A},m\neq n$, it satisfies
\begin{equation}
\frac{\partial f_n^k(\textbf{a}_{-n})}{\partial a_m^{k'}}=\left\{
\begin{array}{cl}
F_{mn}^k,& \text{if $k'=k$}\\
0,& \text{otherwise}.
\end{array} \right.
\end{equation} In addition, we can see that, in Example \ref{ex:ex4}, $f_n^k(\cdot)$ is
actually an affine function with
\begin{equation}
\frac{\partial f_n^k(\textbf{P}_{-n})}{\partial P_m^{k'}}=\left\{
\begin{array}{cl}
\gamma(k-k')H_{mn}^{k'},& \text{if $k'=k$}\\
0,& \text{otherwise}.
\end{array} \right.
\end{equation} and $\bar{\textbf{S}}^{\max}$ is reduced to
\begin{equation}\label{eq:newSbarmax}
[\bar{\textbf{S}}^{\max}]_{mn}\triangleq\left\{ \begin{array}{cl}
\max_{k'}\sum_{k=1}^K\gamma(k-k')H_{mn}^{k'},& \text{if $m\neq n$}\\
0,& \text{otherwise}.
\end{array} \right.
\end{equation}
As an immediate result of Theorem \ref{th:th4}, we have the
following corollary which specifies a sufficient condition that
guarantees the convergence of the iterative water-filling algorithm
for asynchronous transmissions in multi-carrier systems \cite{ASB}.
\end{remark}

\begin{corollary}
In Example \ref{ex:ex4}, if the matrix $\bar{\textbf{S}}^{\max}$
defined in (\ref{eq:newSbarmax}) satisfies
\begin{equation}
\rho(\bar{\textbf{S}}^{\max})<1,
\end{equation}
then there exists a unique NE in game $\Gamma$ and the iterative
water-filling algorithm converges linearly to the NE, for any set of
initial conditions belonging to $\mathcal{A}$ and with either
sequential or parallel updates.
\end{corollary}

\begin{remark}\label{rm:rm4.5}
\textit{(Impact of sum constraints)} An interesting phenomenon that
can be observed from the analysis above is that, the convergence
condition may depend on the maximum constraints $\{M_n\}_{n=1}^N$.
This differs from the observation in \cite{Scutari} that the
presence of the transmit power and spectral mask constraints does
not affect the convergence capability of the iterative water-filling
algorithm. This is because when functions $f_n^k(\textbf{a}_{-n})$
are affine, e.g. in Example \ref{ex:ex2}, \ref{ex:ex3}, and
\ref{ex:ex4}, the elements in $\bar{\textbf{T}}^{\max}$ and
$\bar{\textbf{S}}^{\max}$ are independent of the values of
$\{M_n\}_{n=1}^N$. Therefore, (\ref{C:C1})-(\ref{C:C6}) are
independent of $M_n$ for affine $f_n^k(\textbf{a}_{-n})$. However,
for non-linear $f_n^k(\textbf{a}_{-n})$, the values of
$\{M_n\}_{n=1}^N$ specify the range of users' joint feasible action
set $\mathcal{A}$, and this will affect $\bar{\textbf{T}}^{\max}$
and $\bar{\textbf{S}}^{\max}$ accordingly. In other words, in the
presence of non-linearly coupled $f_n^k(\textbf{a}_{-n})$,
convergence may depend on the players' maximum sum constraints
$\{M_n\}_{n=1}^N$.
\end{remark}

\subsection{Connections to the  Results of Rosen \cite{Rosen} and Gabay\cite{Moulin}}
In \cite{Rosen}, Rosen proposed a continuous-time gradient
projection based iterative algorithm to obtain a pure NE under the
assumption of DSC conditions. Here we present a discrete version of
the algorithm in \cite{Rosen}, named ``\textit{gradient play}''.
Specifically, at stage $t$, each user first determines the gradient
of its own utility function
$u_n(\mathbf{a}_n,\mathbf{a}_{-n}^{t-1})$. Then each user updates
its action $a_n^t$ using gradient projection according to
\begin{equation}
\label{eq:eqngp1}a_n^{'k,t}=a_n^{k,t-1}+\kappa_n\frac{\partial
u_n(\mathbf{a}_n,\mathbf{a}_{-n}^{t-1})}{\partial a_n^{k}}
\end{equation}
and
\begin{equation}
\label{eq:eqngp2}\textbf{a}_n^{t}=[a_n^{1,t} a_n^{2,t} \cdots
a_n^{K,t}]=\Big[a_n^{'1,t} a_n^{'2,t} \cdots
a_n^{'K,t}\Big]_{\mathcal{A}_n}^{\|\cdot\|_2},
\end{equation}
where $\kappa_n$ is the stepsize and
$[\textbf{v}]_{\mathcal{A}_n}^{\|\cdot\|_2}$ denotes the projection
of the vector $\textbf{v}$ onto user $n$'s action set
$\mathcal{A}_n$ with respect to the Euclidean norm $\|\cdot\|_2$. If
$\kappa_n$ is chosen to be sufficiently small, gradient play
approximates the continuous-time gradient projection algorithm. For
each nonnegative vector $\pmb{\kappa}=[\kappa_1 \ldots \kappa_N]$,
define
\begin{equation}
g(\mathbf{a},\pmb{\kappa})=[\kappa_1\nabla_1u_1(\mathbf{a})\
\kappa_2\nabla_2u_2(\mathbf{a})\ \ldots \
\kappa_N\nabla_Nu_N(\mathbf{a})]^T.
\end{equation} The definition of DSC in \cite{Rosen} is that,
for fixed $\pmb{\kappa}>0$ and every $\mathbf{a}^0,\mathbf{a}^1\in
\mathcal{A}$, we have
\begin{equation}
(\mathbf{a}^1-\mathbf{a}^0)^Tg(\mathbf{a}^0,\pmb{\kappa})+
(\mathbf{a}^0-\mathbf{a}^1)^Tg(\mathbf{a}^1,\pmb{\kappa})>0.
\end{equation}
A sufficient condition for DSC is that the symmetric matrix
$G(\mathbf{a},\pmb{\kappa})+G^T(\mathbf{a},\pmb{\kappa})$ be
negative definite for $\mathbf{a}\in \mathcal{A}$, where
$G(\mathbf{a},\pmb{\kappa})$ is the Jacobian with respect to
$\mathbf{a}$ of $g(\mathbf{a},\pmb{\kappa})$.

However, when using gradient play to search for a pure NE, the
stepsize $\kappa_n$ needs to be carefully chosen and set to be
sufficiently small, which usually slows down the rate of
convergence. As an alternative distributed algorithm, for concave
games with $\mathcal{A}_n=\mathcal{R}_+$, $\forall n\in
\mathcal{N}$, Gabay and Moulin provided in \cite{Moulin} a dominance
solvability condition under which best response dynamics globally
converges to a unique NE. Specifically, the dominance solvability
condition is given by
\begin{equation}
-\frac{\partial^2 u_n}{\partial^2 a_n}\geq\sum_{m \neq n
}\Big|\frac{\partial^2 u_n}{\partial a_n\partial
a_m}\Big|.\end{equation} The sufficient conditions provided in this
section and Gabay's dominance solvability condition specify the
convergence conditions of best response dynamics in different
subclasses of concave games. Specifically, our results are developed
for concave games in which every user has a multi-dimensional action
space subject to a single sum-constraint and Gabay's dominance
solvability condition is proposed for concave games with single
dimensional strategy.

\begin{table}
\centering \caption{A summary of various convergence conditions in
concave games.}
\begin{tabular}{|c||c|} \hline
Algorithms & Sufficient conditions and the applicable games\\
\hline \hline
Gradient play & Rosen's DSC conditions for concave games \cite{Rosen}\\
\hline
\multirow{2}{*}{Best response} & Gabay's dominance solvability condition for concave games\\
 &with $\mathcal{A}_n=\mathcal{R}_+$ \cite{Moulin}, conditions (\ref{C:C1})-(\ref{C:C6}) for ACSCG \\
\hline
\end{tabular} \label{tb:tb3}
\end{table}

\subsection{Connections to Linearly Coupled Communication Games}
We investigated in \cite{conj_game} the convergence properties in
certain communication scenarios, namely \emph{linearly coupled
communication games} (LCCG), in which each user has a convex action
set $\mathcal{A}_n\subseteq\mathcal{R}_+$ and the utility functions
take the form
\begin{equation}
u_n(\mathbf{a})=a_n^{\beta_n}\cdot (\mu-\sum_{m=1}^N\tau_ma_m).
\end{equation}
It has been used to model the flow control mechanism in
communication networks \cite{Zhang_TCOM}. In best response dynamics,
at stage $t$, user $n$ chooses its action according to
\begin{equation}
\label{eq:linear_BR}B_n(\textbf{a}^{t-1})= \frac
{\beta_n(\mu-\sum_{m \in
\mathcal{N}\setminus\{n\}}\tau_ma_m^{t-1})}{\tau_n(1+\beta_n)}.
\end{equation}
We can see that, LCCG is similar to ACSCG in the sense that the best
response iterations at stage $t$ in (\ref{eq:eqn9}) and
(\ref{eq:linear_BR}) both contain the linear combinations of
$\textbf{a}^{t-1}$. However, since
$\mathcal{A}_n\subseteq\mathcal{R}$ in LCCG,  we can explicitly
derive the Jacobian matrix for best response dynamics and determine
the exact locations of all its eigenvalues. Consequently, we are
able to develop the necessary and sufficient condition that ensures
the spectral radius of the Jacobian matrix to be less than 1 and
best response dynamics globally converges. However, in ACSCG, due to
the sum-constraint, there exists a non-linear operation $[x]_b^a$ in
equation (\ref{eq:eqn9}). This complicates the analysis of the
Jacobian matrix's eigenvalues. Therefore, we usually choose various
appropriate matrix norms to bound the spectral radius of the
Jacobian matrix and ensure the best response iteration to converge
under these matrix norms. This approach generally results in various
sufficient, but not necessary, conditions.

\section{Scenario II: message exchange among users}
\label{sc:cp}In this section, our objective is to coordinate the
users' actions in ACSCG to maximize the overall performance of the
system, measured in terms of their total utilities, in a distributed
fashion. Specifically, the optimization problem we want to solve is
\begin{equation}\label{eq:eqn25}
\max_{\textbf{a}\in \mathcal{A}}\sum_{n=1}^N u_n(\mathbf{a}).
\end{equation}
We will study two distributed algorithms in which the participating
users exchange price signals that indicate the ``cost'' or
``benefit'' that its action causes to the other users. Allocating
network resources via pricing has been well-investigated for convex
NUM problems \cite{Num}, where the original NUM problem can be
decomposed into distributedly solvable subproblems by setting price
for each constraint resource, and each subproblem has to decide the
amount of resources to be used depending on the charged price.
However, unlike in the conventional convex NUM, pricing mechanisms
may not be immediately applicable in ACSCG if the objective in
(\ref{eq:eqn25}) is not jointly concave in $\mathbf{a}$. Therefore,
we are interested in characterizing the convergence condition of
different pricing algorithms in ACSCG.

We know that for any local maximum $\textbf{a}^*$ of problem
(\ref{eq:eqn25}), there exist Lagrange multipliers $\lambda_n,
\nu^1_n,\cdots,\nu^N_n$ and $\nu'^1_n,\cdots,\nu'^N_n$ such that the
following Karush-Kuhn-Tucker (KKT) conditions hold for all $n \in
\mathcal{N}$:
\begin{align}
\label{eq:eqn26}&\frac{\partial u_n(\mathbf{a}^*)}{\partial a_n^{k}}+\sum_{m\neq n}\frac{\partial u_m(\mathbf{a}^*)}{\partial a_n^{k}}=\lambda_n+\nu^k_n-\nu^{'k}_n,\ \forall n\\
\label{eq:eqn27}&\lambda_n\Big(\sum_{k=1}^Ka_n^{k*}-M_n\Big)=0,\ \lambda_n\geq0\\
\label{eq:eqn28}&\nu^{k}_n(a_n^{k*}-a_{n,k}^{\max})=0, \
\nu^{'k}_n(a_{n,k}^{\min}-a_n^{k*})=0,\ \nu^{k}_n,\nu^{'k}_n\geq0.
\end{align}
Denote $\pi_{mn}^k$ user $m$'s marginal fluctuation in utility per
unit decrease in user $n$'s action $a_n^k$ within the $k$th
dimension
\begin{equation}
\label{eq:eqn29}\pi_{mn}^k(a_m^k,\textbf{a}_{-m}^k)=-\frac{\partial
u_m(\mathbf{a})}{\partial a_n^k},
\end{equation}
which is announced by user $m$ to user $n$ and can be viewed as the
cost charged (or compensation paid) to user $n$ for changing user
$m$'s utility. Using (\ref{eq:eqn29}), equation (\ref{eq:eqn26}) can
be rewritten as
\begin{equation}
\label{eq:eqn30}\frac{\partial u_n(\mathbf{a}^*)}{\partial
a_n^{k}}-\sum_{m\neq
n}\pi_{mn}^k(a_m^{k*},\textbf{a}_{-m}^{k*})=\lambda_n+\nu^{k}_n-\nu^{'k}_n.
\end{equation}
If we assume fixed prices $\{\pi_{mn}^k\}$ and action profile
$\textbf{a}_{-n}^k$, condition (\ref{eq:eqn30}) gives the necessary
and sufficient KKT condition of the following problem:
\begin{equation}
\label{eq:eqn31}\max_{\textbf{a}_n\in\mathcal{A}_n}\
u_n(\mathbf{a})-\sum_{k=1}^Ka_n^k\cdot\Big(\sum_{m\neq
n}\pi_{mn}^k\Big).
\end{equation}
At an optimum, a user behaves as if it maximizes the differences
between its utility minus its payment to the other users in the
network due to its impact over the other users' utilities. Different
distributed pricing mechanisms can be developed based on the
individual objective function in (\ref{eq:eqn31}) and the
convergence conditions may also vary based on the specific action
update equation.

When optimization program (\ref{eq:eqn25}) is not convex, the
pricing algorithms developed for convex NUM, e.g. gradient and
subgradient algorithms, cannot be directly applied. In the next two
subsections, we will investigate two distributed pricing mechanisms
for non-convex ACSCG and provide two sufficient conditions that
guarantee their convergence. Specifically, under these sufficient
conditions, both algorithms guarantee that the total utility is
monotonically increasing until it converges to a feasible operating
point that satisfies the KKT conditions. Similarly as in Section
\ref{sc:l}, we first assume $f_n^k(\mathbf{a}_{-n})$ takes the form
in (\ref{eq:simplef}) and users update their actions in parallel.

\subsection{Gradient Play}
The first distributed pricing algorithm that we consider is gradient
play. The update iterations of gradient play need to be properly
redefined in presence of real-time information exchange.
Specifically, at stage $t$, users adopting this algorithm exchange
price signals $\{\pi_{mn}^{k,t-1}\}$ using the gradient information
at stage $t-1$. Within each iteration, each user first determines
the gradient of the objective in (\ref{eq:eqn31}) based on the price
vectors $\{\pi_{mn}^{k,t-1}\}$ and its own utility function
$u_n(\mathbf{a}_n,\mathbf{a}_{-n}^{t-1})$. Then each user updates
its action $a_n^t$ using gradient projection algorithm according to
\begin{equation}
\label{eq:eqn32} a_n^{'k,t}=a_n^{k,t-1}+\kappa\Big(\frac{\partial
u_n(\mathbf{a}_n,\mathbf{a}_{-n}^{t-1})}{\partial
a_n^{k}}-\sum_{m\neq n}\pi_{mn}^{k,t-1}\Big).
\end{equation}
and
\begin{equation}
\label{eq:eqn33} \textbf{a}_n^{t}=[a_n^{1,t} a_n^{2,t} \cdots
a_n^{K,t}]=\Big[a_n^{'1,t} a_n^{'2,t} \cdots
a_n^{'K,t}\Big]_{\mathcal{A}_n}^{\|\cdot\|_2}.
\end{equation}
in which the stepsize $\kappa>0$. The following theorem provides a
sufficient condition under which gradient play will converge
monotonically provided that we choose small enough constant stepsize
$\kappa$.

\begin{theorem}\label{th:th5}
If $\forall n,k,\mathbf{x},\mathbf{y}\in\mathcal{A}_{-n}$,
\begin{equation}
\label{C:C7} \tag{C7} \inf_{x}\frac{\partial^2
h_n^{k}(x)}{\partial^2 x}>-\infty, \ \textrm{and} \
\Big\|\bigtriangledown g_n^k(\mathbf{x})-\bigtriangledown
g_n^k(\mathbf{y})\Big\|\leq L'\big\|\mathbf{x}-\mathbf{y}\big\|,
\end{equation} gradient play converges for a small
enough stepsize $\kappa$.
%\begin{equation}
%0<\kappa<\frac{2}{\sup_{x,n}|h''_n(x)|\cdot\max_{k,l}\sum_{m=1}^N\sum_{n=1}^N|F_{mn}^kF_{ln}^k|+NKL'}.
%\end{equation}
\end{theorem}

\textit{Proof}: This theorem can be proved by showing the gradient
of the objective function in (\ref{eq:eqn25}) is Lipschitz
continuous and applying Proposition 3.4 in \cite{Bertsekas}. See
Appendix E for details. $\blacksquare$

\begin{remark}\label{rm:rm5}
\textit{(Application of condition (\ref{C:C7}))} A sufficient
condition that guarantees the convergence of distributed gradient
projection algorithm is the Lipschitz continuity of the gradient of
the objective function in (\ref{eq:eqn25}). For example, in the
power control problem in multi-channel networks \cite{Huang_JSAC},
we have $h_n^k(x)=\log_2(\alpha_n^k+H_{nn}^kx)$ and
$g_n^k(\textbf{P}_{-n})=\log_2(\sigma_n^k+\sum_{m\neq n}
H_{mn}^kP_m^k)$. For this configuration, we can immediately verify
that condition (\ref{C:C7}) is satisfied. Therefore, gradient play
can be applied. Moreover, as in \cite{Huang_JSAC}, if we can further
ensure that the problem in (\ref{eq:eqn25}) is convex for some
particular utility functions, gradient play converges to the unique
optimal solution of (\ref{eq:eqn25}) at which achieving KKT
conditions implies global optimality.
\end{remark}

%\begin{remark}\label{rm:rm5}
%\textit{(Relation between conditions (\ref{C:C7}) and the result in
%\cite{Huang_JSAC})} The key property that guarantees the convergence
%of distributed gradient projection algorithm is actually the
%Lipschitz continuity of the gradient of the objective function in
%(\ref{eq:eqn25}). Condition (\ref{C:C7}) extends the convergence
%condition of the Dual Asynchronous Distributed Pricing (ADP)
%algorithm in \cite{Huang_JSAC} developed for power control in
%multi-channel networks to general ACSCG. In particular, in
%\cite{Huang_JSAC}, we have $h_n^k(x)=-g_n^k(x)=\log_2(x), \forall
%n,k$, $a_{n,k}^{\min}>0$, and $a_{n,k}^{\max}<\infty$. For this
%configuration, we can immediately verify that condition (\ref{C:C7})
%is satisfied. Moreover, as in \cite{Huang_JSAC}, if we can further
%ensure that the problem in (\ref{eq:eqn25}) is convex based on the
%specific utility forms, gradient play algorithm converges to the
%unique optimal solution of (\ref{eq:eqn25}) at which achieving KKT
%conditions implies global optimality.
%\end{remark}

\subsection{Jacobi Update}
We consider another alternative strategy update mechanism called
Jacobi update \cite{Jacobi_play}. In Jacobi update, every user
adjusts its action gradually towards the best response strategy.
Specifically, the maximizer of problem (\ref{eq:eqn31}) takes the
following form
\begin{equation} \label{eq:eqn34}
B_n^{'k}(\mathbf{a}_{-n})= \Big\{\frac{\partial h^{k}_n}{\partial
x}\Big\}^{-1}\big(\lambda_n+\nu^{k}_n-\nu^{'k}_n+\sum_{m\neq
n}\pi_{mn}^k\big)-\sum_{m\neq n}F_{mn}^ka_m^k,
\end{equation}
in which $\lambda_n$, $\nu^{k}_n$, and $\nu^{'k}_n$ are the Lagrange
multipliers that satisfy complementary slackness in (\ref{eq:eqn27})
and (\ref{eq:eqn28}), and $\pi_{mn}^k$ is defined in
(\ref{eq:eqn29}). In Jacobi update, at stage $t$, user $n$ chooses
its action according to
\begin{equation}
\label{eq:eqn35} a_n^{k,t}=a_n^{k,t-1}+
\kappa\bigl[B_n^{'k}(\textbf{a}_{-n}^{t-1})-a_n^{k,t-1}\bigr],
\end{equation}
in which the stepsize $\kappa\in (0,1]$. The following theorem
establishes a sufficient convergence condition for Jacobi update.
\begin{theorem}
\label{th:th6} If $\forall
n,k,\mathbf{x},\mathbf{y}\in\mathcal{A}_{-n}$,
\begin{equation}
\label{C:C8} \tag{C8} \inf_{x}\frac{\partial^2
h_n^{k}(x)}{\partial^2 x}>-\infty, \ \sup_{x}\frac{\partial^2
h_n^{k}(x)}{\partial^2 x}<0, \ \textrm{and} \ \Big\|\bigtriangledown
g_n^k(\mathbf{x})-\bigtriangledown g_n^k(\mathbf{y})\Big\|\leq
L'\big\|\mathbf{x}-\mathbf{y}\big\|,
\end{equation}Jacobi update converges if the
stepsize $\kappa$ is sufficiently small.
\end{theorem}

\emph{Proof}: This can be proved using the descent lemma and the
mean value theorem. The details of the proof are provided in
Appendix F. $\blacksquare$

\begin{remark}\label{rm:rm6}
\textit{(Relation between condition (\ref{C:C8}) and the result in
\cite{Shi})} Shi et al. considered the power allocation for
multi-carrier wireless networks with non-separable utilities.
Specifically, $u_n(\cdot)$ takes the form
\begin{equation}
u_n(\mathbf{P})=r_i\Bigg(\sum_{k=1}^K\log_2\Big(1+\frac{H_{nn}^kP_n^k}{\sigma_n^k+\sum_{m\neq
n} H_{mn}^kP_m^k}\Big)\Bigg),
\end{equation}
in which $r_i(\cdot)$ is an increasing and strictly concave
function. Since the utilities are non-separable, the distributed
pricing algorithm proposed in \cite{Shi}, which in fact belongs to
Jacobi update, requires only one user to update its action profile
at each stage while keeping the remaining users' action fixed. The
condition in (\ref{C:C8}) gives the convergence condition of the
same algorithm in ACSCG. We prove in Theorem \ref{th:th5} that, if
the utilities are separable, convergence can still be achieved even
if these users update their actions at the same time. Therefore, we
do not need an arbitrator to select the single user that updates its
action at each stage.
\end{remark}

\begin{remark}\label{rm:rm7}
\textit{(Complexity of signaling)} The complexity of message
exchange measured in terms of the number of price signals to update
in (\ref{eq:eqn29}) is generally of the order of $O(KN^2)$. It is
worth mentioning that the amount of signaling can be further reduced
to $O(KN)$ in the scenarios where $g_n^k(\cdot)$ are functions of
$\sum_{m\neq n}F_{mn}^ka_m^k$. In this case, each user only needs to
announce one price signal $\pi_{n}^k$ for each dimension of its
action space:
\begin{equation}
\pi_{n}^k(a_n^k,\textbf{a}_{-n}^k)=-\frac{\partial
u_n(\mathbf{a})}{\partial \big(\sum_{m\neq n}F_{mn}^ka_m^k\big)}
\end{equation}
Consequently, $\pi_{mn}^k$ can be determined based on
$\pi_{mn}^k=F_{nm}^k\pi_{m}^k$, which greatly reduces the overhead
of signaling requirement. It is straightforward to check that only
$O(KN)$ messages need to be generated and exchanged per iteration in
both utility functions (\ref{eq:eqn3}) and (\ref{eq:eqn4}).
\end{remark}

\begin{remark}\label{rm:rm8}
\textit{(Extension to general cases)} As a matter of fact,
conditions (\ref{C:C7}) and (\ref{C:C8}) apply to a broader class of
multi-user interaction scenarios, including the general model
defined in (\ref{eq:eqn2}). Specifically, as addressed in Remark
\ref{rm:rm5}, the Lipschitz continuity of the gradient of
$\sum_{n=1}^N u_n(\mathbf{a})$ is sufficient to guarantee that
gradient play with a small enough stepsize achieves an operating
point at which KKT conditions are satisfied. In addition, we can use
the same technique in Appendix F to show the convergence of Jacobi
update given that $\sup_{x}\frac{\partial^2 h_n^{k}(x)}{\partial^2
x}<0$, $\forall n,k$, and the gradient of $\sum_{n=1}^N
u_n(\mathbf{a})$ is Lipschitz continuous.
\end{remark}

\section{Numerical Examples}
In Section \ref{sc:ex}, we present several illustrative examples of
ACSCG. This section uses Examples \ref{ex:ex1} and \ref{ex:ex3} to
illustrate the various distributed algorithms discussed in the
paper.

\begin{figure}
\centering
\includegraphics[width=0.6\textwidth]{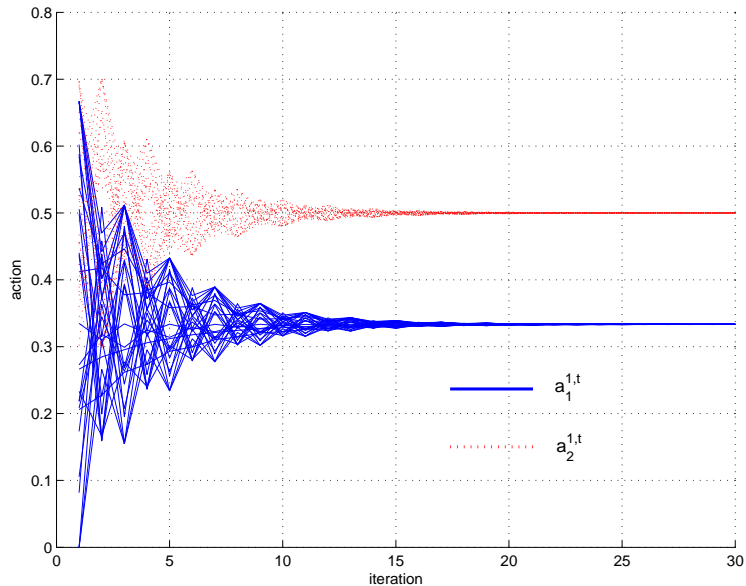}
\caption{Actions versus iterations in Example \ref{ex:ex1}.}
\label{fg:fg1.5}
\end{figure}

We start with Example \ref{ex:ex1} to verify the proposed
convergence conditions of best response dynamics. Even though it is
a simple two-user game with $\mathcal{A}_n\subseteq \mathcal{R}^2$,
existing results in the literature cannot immediately determine
whether or not the best response dynamics in this simple game can
globally converge to a NE. Specifically, in Example \ref{ex:ex1}, we
have
\begin{equation}
\begin{gathered}
\frac{\partial f_n^1(\textbf{a}_{-n})}{\partial
a_{-n}^{1}}=\frac{a_{-n}^{1}}{\sqrt{(a_{-n}^{1})^2+1}},
\frac{\partial f_n^1(\textbf{a}_{-n})}{\partial
a_{-n}^{2}}=-\frac{a_{-n}^{2}}{\sqrt{(a_{-n}^{2})^2+1}},\\
\frac{\partial f_n^2(\textbf{a}_{-n})}{\partial
a_{-n}^{1}}=-\frac{a_{-n}^{1}}{\sqrt{(a_{-n}^{1})^2+1}},
\frac{\partial f_n^2(\textbf{a}_{-n})}{\partial
a_{-n}^{2}}=\frac{a_{-n}^{2}}{\sqrt{(a_{-n}^{2})^2+1}}.
\end{gathered}
\end{equation}
According the definition of (\ref{eq:Tbarmax}), we have
\begin{equation}
\begin{array}{l}
[\bar{\textbf{T}}^{\max}]_{12}=
\max\Big\{\max_{\textbf{a}\in\mathcal{A}}\sum_{k=1}^K\Big|\frac{\partial
f_2^k(\textbf{a}_{-n})}{\partial
a_1^1}\Big|,\max_{\textbf{a}\in\mathcal{A}}\sum_{k=1}^K\Big|\frac{\partial
f_2^k(\textbf{a}_{-n})}{\partial a_1^2}\Big|\Big\}\\
=\max\Big\{\max_{\textbf{a}_1\in\mathcal{A}_1}\frac{2a_{1}^{1}}{\sqrt{(a_{1}^{1})^2+1}},
\max_{\textbf{a}_1\in\mathcal{A}_1}\frac{2a_{1}^{2}}{\sqrt{(a_{1}^{2})^2+1}}\Big\}=\frac{2M_1}{\sqrt{M_1^2+1}}.
\end{array}\end{equation}
Similarly, we can obtain
$[\bar{\textbf{T}}^{\max}]_{21}=\frac{2M_2}{\sqrt{M_2^2+1}}$.
Therefore,
$\rho(\bar{\textbf{T}}^{\max})=\sqrt{\frac{4M_1M_2}{\sqrt{M_1^2+1}\sqrt{M_2^2+1}}}$.
It is easy to show that
$\rho(\bar{\textbf{T}}^{\max})<1\Leftrightarrow
(M_1^2-\frac{1}{3})(M_1^2-\frac{1}{3})<\frac{4}{9}$. By condition
(\ref{C:C4}), we know that if
$(M_1^2-\frac{1}{3})(M_1^2-\frac{1}{3})<\frac{4}{9}$, the best
response dynamics is guaranteed to converge to a unique NE. We
numerically simulate a scenario with parameters $M_1=\frac{2}{3}$
and $M_2=1$ in which condition (\ref{C:C4}) holds. We generate
multiple initial action profiles of $\textbf{a}_1^0$ and
$\textbf{a}_2^0$, iterate the best response dynamics, and obtain the
action sequences $\textbf{a}_1^t$ and $\textbf{a}_2^t$. Fig.
\ref{fg:fg1.5} shows the trajectories of $a_1^{1,t}$ and $a_2^{1,t}$
for different realizations. We can see that, best response dynamics
converges to a unique NE. If we set $M_1=2$ and $M_2=1$, condition
(\ref{C:C4}) does not hold any more. We observe from simulations
that in many circumstances the best response dynamics will not
converge, which agrees with our analysis in Remark \ref{rm:rm4.5}.

Now we consider Example \ref{ex:ex3}, which is the problem of
minimizing queueing delays in a Jackson network. In particular, we
consider a network with $N=5$ nodes and $K=3$ traffic classes. The
total routing probability $1-r_{m0}^k$ that node $m$ will route
packets of class $k$ completing service to other nodes is the same
for $\forall m\in\mathcal{N}$. We varied the total routing
probability $1-r_{m0}^k$ and generated multiple sets of network
parameters in which $r_{mn}^k$ are uniformly distributed for
$n=1,2,\cdots,N$, $\mu_{n}^k$ are uniformly selected in $[4,5]$ for
$\forall n,k$, and $\psi_{n}^{\min}$ are uniformly chosen in
$[0.6,1]$ for $n=1,2,\cdots,N$.

\begin{figure}
\centering
\includegraphics[width=0.6\textwidth]{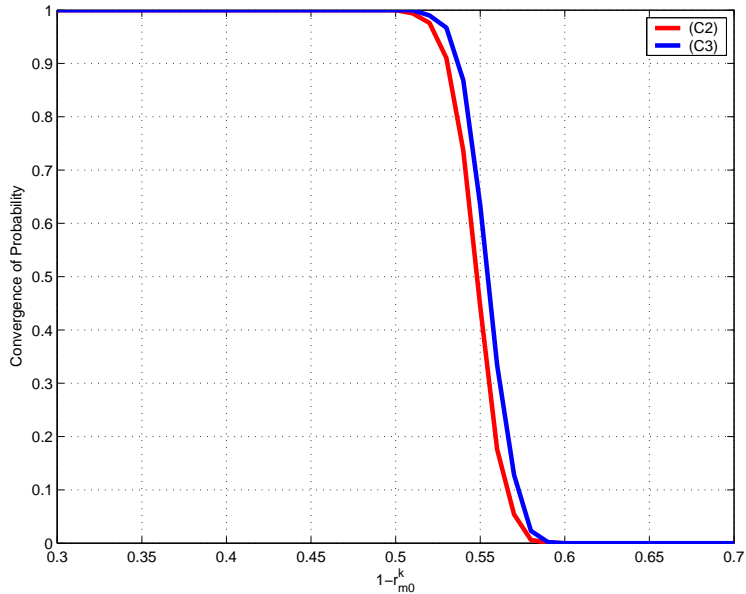}
\caption{Probability of (\ref{C:C2}) and (\ref{C:C3}) versus
$1-r_{m0}^k$ for $\forall m,k$, $N=5$, $K=3$.} \label{fg:fg2}
\end{figure}

\begin{figure}
\centering
\includegraphics[width=0.6\textwidth]{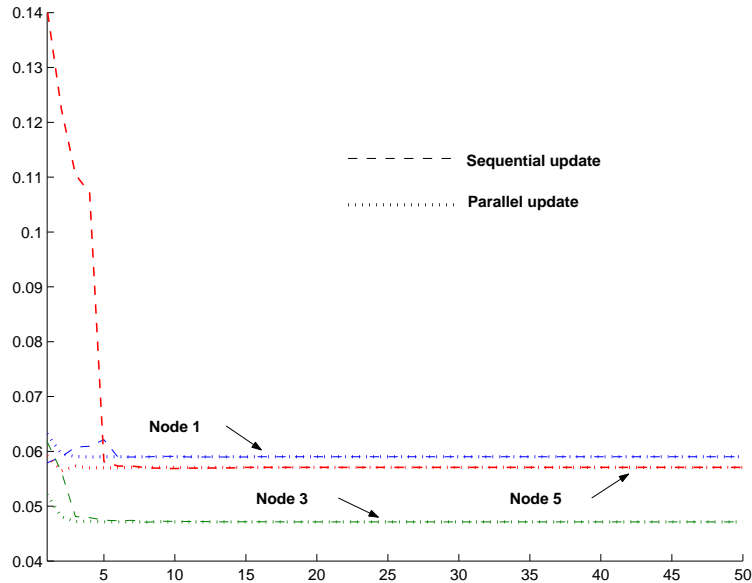}
\caption{Delays of nodes versus iterations.} \label{fg:fg4}
\end{figure}

First of all, we compare the range of validity of the proposed
convergence conditions. As we mentioned before, we have
$F_{mn}^k=\frac{[(\textrm{I}-\textrm{R}^k)^{-1}]_{nm}}{[(\textrm{I}-\textrm{R}^k)^{-1}]_{nn}}$
in this example. Note that
$(\textrm{I}-\textrm{R}^k)^{-1}=\textrm{I}+\sum_{i=1}^{\infty}(\textrm{R}^k)^i$
and $\textrm{R}^k$ is a non-negative matrix. Therefore, we can
conclude $F_{mn}^k\geq0, \forall m\neq n,k$. Moreover, since
$h^k_n(x)=-\frac{1}{\mu_n^k-\upsilon_{nn}^kx}$, we choose to compare
conditions (\ref{C:C2}) and (\ref{C:C3}). In Fig. \ref{fg:fg2}, we
plot the probability that conditions (\ref{C:C2}) and (\ref{C:C3})
are satisfied versus the total routing probability $1-r_{m0}^k$.
From Fig. \ref{fg:fg2}, we can see that the probability of
guaranteeing convergence decreases as the routing probability
$1-r_{m0}^k$ increases and condition (\ref{C:C3}) shows a similar
but slightly broader validity than (\ref{C:C2}). Fig. \ref{fg:fg4}
shows the delay trajectories of three nodes using both sequential
and parallel updates in a certain network realization in which
(\ref{C:C2}) and (\ref{C:C3}) are satisfied. We can see that, the
parallel update converges faster than the sequential update.

In Fig. \ref{fg:fg2}, we also note that the probability that
(\ref{C:C2}) or (\ref{C:C3}) is satisfied transits very quickly from
the almost certain convergence to the non-convergence guarantee as
$1-r_{m0}^k$ varies from $0.5$ to $0.58$. Similar observations have
been drawn in the multi-channel power control problem
\cite{Scutari}, where $\theta=-1$ in (\ref{eq:eqn14}) and the
probability that condition (\ref{C:C3}) is satisfied exhibits a neat
threshold behavior as the ratio between the source-interferer
distance and the source-destination distance varies. In Jackson
networks, this threshold can be roughly estimated. Define
$[\textbf{S}^k]_{mn}=F^k_{mn}$ for $m\neq n$ and
$[\textbf{S}^k]_{nn}=0$ for $n\in \mathcal{N}$. If we fix
$1-r_{m0}^k$ for $\forall m,k$, we prove in Appendix G that
$\rho(\textbf{S}^k)\leq\frac{1}{r_{m0}^k}-1$ for $\forall k$.
Therefore, $\rho(\textbf{S}^k)<1$ when $r_{m0}^k>0.5$. We would like
to estimate $\rho(\textbf{T}^{\max})$ and $\rho(\textbf{S}^{\max})$
based on $\rho(\textbf{S}^k)$. Note that $\textbf{T}^{\max}$ defined
in (\ref{eq:Tmax}) is the element-wise maximum over $\textbf{S}^k$
for $k=1,2,\ldots,K$. Since $\textbf{T}^{\max}$ and $\textbf{S}^k$
are all non-negative matrices, we know that
$\rho(\textbf{T}^{\max})\geq\max_{k}\rho(\textbf{S}^k)$. In
addition, recall the effect of $\max_{m,n}\zeta_{mn}$ discussed in
Remark \ref{rm:rm3}. We can approximate $\rho(\textbf{S}^{\max})$
defined in (\ref{eq:eqn16}) using $\rho(\textbf{S}^{\max})\approx
\max_{m,n}\zeta_{mn}\max_{k}\rho(\textbf{S}^k)$. Therefore, we
expect that $\rho(\textbf{T}^{\max})$ and $\rho(\textbf{S}^{\max})$
exceeds 1 for $r_{m0}^k<0.5$, which agrees with our observation from
Fig. \ref{fg:fg2}. The physical interpretation is that, if the
packets exit the network with a probability less than $50\%$ after
completing its service, i.e. more than half of the served packets
will be routed to other nodes, the strength of the mutual coupling
among users becomes too strong and the multi-user interaction in
Jackson networks will gradually lose its convergence guarantee.

\begin{figure}
\centering
\includegraphics[width=0.6\textwidth]{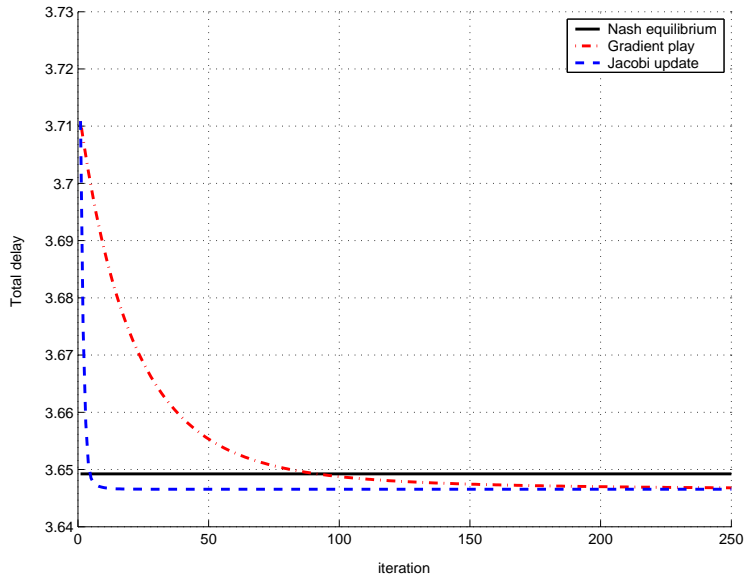}
\caption{Illustration of convergence for gradient play and Jacobi
update.} \label{fg:fg3}
\end{figure}

In addition, we numerically compare two distributed algorithms in
which users pass coordination messages in real time, including
Jacobi update and gradient play. Fig. \ref{fg:fg3} shows the delay
evolution of both distributed solutions for a particular simulated
network in which we set $\kappa=0.2$. We initialize the system
parameters such that
$\inf_{n,k}\mu_n^k-\sum_{m=1}^N\upsilon_{mn}^k\psi_m^k>0$ and both
conditions (\ref{C:C7}) and (\ref{C:C8}) are satisfied. We can
verify that for Example \ref{ex:ex3}, problem (\ref{eq:eqn25}) is in
fact a convex program. Therefore, there exists a unique operating
point at which KKT conditions (\ref{eq:eqn26})-(\ref{eq:eqn28}) are
satisfied. We can see that, both algorithms cause the total delay to
monotonically decrease until it reaches the same performance limit
that is strictly better than NE. Using the same stepsize $\kappa$,
Jacobi update converges more quickly than gradient play in this
example. Similar observations are drawn in the other simulated
examples. This is because the update directions of these two
algorithms are different. Jacobi update moves directly towards the
optimal solution of (\ref{eq:eqn31}), which is a local approximation
of the original optimization program in (\ref{eq:eqn25}), whereas
the gradient play algorithm simply updates the actions along the
gradient direction of (\ref{eq:eqn25}).

\section{Conclusion}
In this paper, we propose and investigate a new game model, which we
refer to as additively coupled sum constrained games, in which each
player is subject to a sum constraint and its utility is additively
impacted by the remaining users' actions. The convergence properties
of various generic distributed adjustment algorithms, including best
response, gradient play, and Jacobi update, have been investigated.
The sufficient conditions obtained in this paper generalize the
existing results developed in the multi-channel power control
problem and can be extended to other applications that belong to
ACSCG.

\appendices
\section{Proof of Theorem \ref{th:th1}}
The following lemma is needed to prove Theorem \ref{th:th1}.
\begin{lemma}\label{lm:lm1}
Consider any non-decreasing function $p(x)$ and non-increasing
function $q(x)$. If there exists a unique $x^*$ such that
$p(x^*)=q(x^*)$, and the functions $p(x)$ and $q(x)$ are strictly
increasing and strictly decreasing at $x=x^*$ respectively, then
$x^*=\arg\min_x\{\max\{p(x),q(x)\}\}$.
\end{lemma}

\textit{Proof of Lemma \ref{lm:lm1}:} See Lemma 1 in \cite{ASB}.
$\blacksquare$

Denote $a_n^{k,t}$ as the action of user $n$ in the $k$th dimension
after iteration $t$. Recall that $\bigl[h_n^k\bigr]'(\cdot)>0$,
for $\forall n,k$. Therefore, $\sum_{k=1}^Ka_n^{k,t}=M_n$ is
satisfied at the end of any iteration $t$ for any user $n$. Define
$[x]^+=\max\{x,0\}$ and $[x]^-=\max\{-x,0\}$. It is straightforward
to see that
\begin{equation}
\label{eq:eqnA1}
\sum_{k=1}^K[a_n^{k,t}-a_n^{k,t'}]^+=\sum_{k=1}^K[a_n^{k,t}-a_n^{k,t'}]^-,
\forall n,t,t'.
\end{equation}
We also define
\begin{equation}
\label{eq:eqnA2}
p^{n,t}(x)\triangleq\sum_{k=1}^K\Big[l_n^k(\mathbf{a}^{t}_{-n},x)-a_n^{k,t}\Big]^{-}
\end{equation}
and
\begin{equation}
\label{eq:eqnA3}
q^{n,t}(x)\triangleq\sum_{k=1}^K\Big[l_n^k(\mathbf{a}^{t}_{-n},x)-a_n^{k,t}\Big]^{+},
\end{equation}
in which $l_n^k(\cdot)$ is defined in (\ref{eq:eqn9}). Since
$h_n^k(\cdot)$ is a continuous increasing and strictly concave
function, it is clear that $\bigl\{\frac{\partial h_n^{k}}{\partial
x}\bigr\}^{-1}(\cdot)$ is a continuous decreasing function. If
$p^{n,t}(\lambda_n^{t+1})\neq0$ (i.e. it has not converged),
$p^{n,t}(x)$ ($q^{n,t}(x)$, respectively) is non-decreasing
(non-increasing) in $x$, and strictly increasing (strictly
decreasing) at $x=\lambda_n^{t+1}$. From (\ref{eq:eqnA1}) it is
always true that
$p^{n,t}(\lambda_n^{t+1})=q^{n,t}(\lambda_n^{t+1})$. We first prove
the convergence of the parallel update case in (\ref{eq:eqn12}). For
$\forall n$, we have
\begin{align}
&\sum_{k=1}^K[a_n^{k,t+1}-a_n^{k,t}]^+\nonumber\\
\label{eq:g0}&=\max\Big\{\sum_{k=1}^K[a_n^{k,t+1}-a_n^{k,t}]^+,\sum_{k=1}^K[a_n^{k,t+1}-a_n^{k,t}]^-\Big\}\\
\label{eq:g1}&=\max\{p^{n,t}(\lambda_n^{t+1}),q^{n,t}(\lambda_n^{t+1})\}\\
\label{eq:g2}&\leq\max\{p^{n,t}(\lambda_n^{t}),q^{n,t}(\lambda_n^{t})\}\\
\label{eq:g3}&\leq\max\Big\{\sum_{k=1}^K\Big[\sum_{m \neq n}F_{mn}^k(a_m^{k,t}-a_m^{k,t-1})\Big]^+,\sum_{k=1}^K\Big[\sum_{m \neq n}F_{mn}^k(a_m^{k,t}-a_m^{k,t-1})\Big]^-\Big\}\\
\label{eq:g3.1}&\leq\max\Big\{\sum_{k=1}^K\sum_{m \neq n}\Big[F_{mn}^k(a_m^{k,t}-a_m^{k,t-1})\Big]^+,\sum_{k=1}^K\sum_{m \neq n}\Big[F_{mn}^k(a_m^{k,t}-a_m^{k,t-1})\Big]^-\Big\}\\
\label{eq:g3.2}&=\max\Big\{\sum_{m \neq n}\sum_{k=1}^K\Big[F_{mn}^k(a_m^{k,t}-a_m^{k,t-1})\Big]^+,\sum_{m \neq n}\sum_{k=1}^K\Big[F_{mn}^k(a_m^{k,t}-a_m^{k,t-1})\Big]^-\Big\}\\
\label{eq:g4}&\leq\sum_{m \neq n}\max_k|F_{mn}^k|\cdot\Big\{\sum_{k=1}^K\Big[a_m^{k,t}-a_m^{k,t-1}\Big]^++\sum_{k=1}^K\Big[a_m^{k,t}-a_m^{k,t-1}\Big]^-\Big\}\\
\label{eq:g5}&=\sum_{m \neq n}2\max_k|F_{mn}^k|\cdot\sum_{k=1}^K\Big[a_m^{k,t}-a_m^{k,t-1}\Big]^+,
\end{align}
where (\ref{eq:g0}) and (\ref{eq:g5}) follows from (\ref{eq:eqnA1}),
(\ref{eq:g1}) follows from the definition of $p^{n,t}$ and $q^{n,t}$
in (\ref{eq:eqnA2}) and (\ref{eq:eqnA3}), (\ref{eq:g2}) is due to
Lemma 1 in which $x=\lambda_n^{t}$, (\ref{eq:g3}) follows from the
definition of $p^{n,t}$ and $q^{n,t}$, the expression of $a_n^{k,t}$
in (\ref{eq:eqn12}), and the fact that
$\big[[x]_a^b-[y]_a^b\big]^+\leq[x-y]^+$ and
$\big[[x]_a^b-[y]_a^b\big]^-\leq[x-y]^-$, (\ref{eq:g3.1}) is due to
the fact that $[x+y]^+\leq[x]^++[y]^+$ and $[x+y]^-\leq[x]^-+[y]^-$,
(\ref{eq:g4}) follows by using $[\sum_kx_ky_k]^+\leq\sum_k|x_k||y_k|
=\sum_k|x_k|([y_k]^++[y_k]^-)\leq\max_k|x_k|\sum_k([y_k]^++[y_k]^-)$.
For user $n$, we define that $e_n^t=\Big[a_m^{k,t}-a_m^{k,t-1}\Big]^+$.
Inequality (\ref{eq:g5}) can be written as $e_n^{t+1}\leq\sum
_{m\neq n}[\textbf{T}^{\max}]_{mn}e_m^t$ in which $\textbf{T}^{\max}$
is defined in (\ref{eq:Tmax}).

Since $\textbf{T}^{\max}$ is a nonnegative matrix, by the
Perron-Frobenius Theorem \cite{Bertsekas}, there exists a positive
vector $\bar{\textbf{w}}=[\bar{w}_1 \ldots \bar{w}_N]$ such that
\begin{equation}
\|\textbf{T}^{\max}\|_{\infty,\textrm{mat}}^{\bar{\textbf{w}}}=\rho(\textbf{T}^{\max}),
\end{equation}
where $\|\cdot\|_{\infty,\textrm{mat}}^{\bar{\textbf{w}}}$ is the
weighted maximum matrix norm defined as
\begin{equation}
\label{eq:specialnorm}
\|\textbf{A}\|_{\infty,\textrm{mat}}^{\bar{\textbf{w}}}\triangleq\max_{i=1,2,\cdots,N}\frac{1}{\bar{w}_i}\sum_{j=1}^N[\textbf{A}]_{ij}\bar{w}_j,
\quad \textbf{A}\in \mathcal{R}^{N\times N}.
\end{equation}

Define the vectors $\textbf{e}^{t+1}\triangleq [e_1^{t+1}, e_2^{t+1}, \ldots,e_N^{t+1}]^T$
and $\textbf{e}^t\triangleq [e_1^t, e_2^t, \ldots,e_N^t]^T$. The set of
inequalities in (\ref{eq:g5}) can be expressed in the vector
form as $\textbf{0} \leq \textbf{e}^{t+1}\leq
\textbf{T}^{\max}\textbf{e}^t$. By choosing the vector
$\bar{\textbf{w}}$ that satisfies
$\|\textbf{T}^{\max}\|_{\infty,\textrm{mat}}^{\bar{\textbf{w}}}=\rho(\textbf{T}^{\max})$
and applying the infinity norm
$\|\cdot\|_{\infty}^{\bar{\textbf{w}}}$, we obtain the following
\begin{equation}
\label{eq:eqnA11}
\|\textbf{e}^{t+1}\|_{\infty}^{\bar{\textbf{w}}}\leq
2\|\textbf{T}^{\max}\textbf{e}^t\|_{\infty}^{\bar{\textbf{w}}}\leq
2\|\textbf{T}^{\max}\|_{\infty,\textrm{mat}}^{\bar{\textbf{w}}}\|\textbf{e}^t\|_{\infty}^{\bar{\textbf{w}}},
\end{equation}
Finally, based on (\ref{eq:g5}) and (\ref{eq:eqnA11}), it
follows that
\begin{align}
\max_{n\in\mathcal{N}}\frac{e_n^{t+1}}{\bar{w}_n}=\|\textbf{e}^{t+1}\|_{\infty}^{\bar{\textbf{w}}}\leq2\|\textbf{T}^{\max}\|_{\infty,\textrm{mat}}^{\bar{\textbf{w}}}\|\textbf{e}^t\|_{\infty}^{\bar{\textbf{w}}}
\leq2\|\textbf{T}^{\max}\|_{\infty,\textrm{mat}}^{\bar{\textbf{w}}}\cdot\max_{n\in\mathcal{N}}\frac{e_n^t}{\bar{w}_n}
=2\rho(\textbf{T}^{\max})\cdot\max_{n\in\mathcal{N}}\frac{e_n^t}{\bar{w}_n}
\end{align}
Therefore, if
$\|\textbf{T}^{\max}\|_{\infty,\textrm{mat}}^{\bar{\textbf{w}}}=\rho(\textbf{T}^{\max})<\frac{1}{2}$,
the best response dynamics in (\ref{eq:eqn12}) is a contraction with the modulus
$\|\textbf{T}^{\max}\|_{\infty,\textrm{mat}}^{\bar{\textbf{w}}}$
with respect to the norm
$\max_{n\in\mathcal{N}}\frac{\|\cdot\|^{\textbf{w}_n}_2}{\bar{w}_n}$.
We can conclude that, the best response dynamics has a
unique fixed point $\mathbf{a}^*$ and, given any initial value
$\mathbf{a}^0$, the update sequence $\bigl\{\mathbf{a}^t\bigr\}$ converges to the fixed point
$\mathbf{a}^*$.

%This shows that, if
%$\max_{m\neq n,k}\frac{|F_{mn}^k|}{F_{nn}^k}<\frac{1}{2(N-1)}$, best
%response dynamics is a contraction map with the modulus
%$\delta=\max_{m\neq n,k}\frac{|F_{mn}^k|}{F_{nn}^k}\cdot2(N-1)<1$
%with respect to the norm $\max_n\{\sum_{k}[x_n^{k}]^+\}$ and it
%globally converges to a unique fixed point.

In the sequential update case, the convergence result can be
established by using the proposition 1.4 in \cite{Bertsekas}. The
key step is to obtain
\begin{align}
\max_{n\in\mathcal{N}}\frac{e_n^{t+1}}{\bar{w}_n}\leq2\rho(\textbf{T}^{\max})\cdot\max\Big\{\max_{j<n}\frac{e_j^{t+1}}{\bar{w}_j},\max_{j\geq
n}\frac{e_j^{t}}{\bar{w}_j}\Big\}.
\end{align}
A simple induction on $n$ yields
\begin{equation}
\max_{n\in\mathcal{N}}\frac{e_n^{t+1}}{\bar{w}_n}\leq2\rho(\textbf{T}^{\max})\cdot
\max_{n\in\mathcal{N}}\frac{e_n^{t}}{\bar{w}_n}
\end{equation}
for all $n$. Therefore, inequality (\ref{eq:g5}) also holds for the
sequential update and the contraction iteration globally converges
to a unique equilibrium. $\blacksquare$

\section{Proof of Theorem \ref{co:co1}}
If $F_{mn}^k\geq0, \forall m\neq n,k$, the inequalities after
(\ref{eq:g3.2}) become
\begin{align}
&\max\Big\{\sum_{m \neq n}\sum_{k=1}^K\Big[F_{mn}^k(a_m^{k,t}-a_m^{k,t-1})\Big]^+,\sum_{m \neq n}\sum_{k=1}^K\Big[F_{mn}^k(a_m^{k,t}-a_m^{k,t-1})\Big]^-\Big\}\\
&\leq\sum_{m \neq n}\max_kF_{mn}^k\cdot\max\Big\{\sum_{k=1}^K\Big[a_m^{k,t}-a_m^{k,t-1}\Big]^+,\sum_{k=1}^K\Big[a_m^{k,t}-a_m^{k,t-1}\Big]^-\Big\}\\
&=\sum_{m \neq n}\max_kF_{mn}^k\cdot\sum_{k=1}^K\Big[a_m^{k,t}-a_m^{k,t-1}\Big]^+.
\end{align}
Similarly, for $F_{mn}^k\leq0, \forall m\neq n,k$, we have
\begin{align}
&\max\Big\{\sum_{m \neq n}\sum_{k=1}^K\Big[F_{mn}^k(a_m^{k,t}-a_m^{k,t-1})\Big]^+,\sum_{m \neq n}\sum_{k=1}^K\Big[F_{mn}^k(a_m^{k,t}-a_m^{k,t-1})\Big]^-\Big\}\\
&\leq\sum_{m \neq n}\max_k\bigl\{-F_{mn}^k\bigr\}\cdot\max\Big\{\sum_{k=1}^K\Big[a_m^{k,t}-a_m^{k,t-1}\Big]^+,\sum_{k=1}^K\Big[a_m^{k,t}-a_m^{k,t-1}\Big]^-\Big\}\\
&=\sum_{m \neq n}\max_k\bigl\{-F_{mn}^k\bigr\}\cdot\sum_{k=1}^K\Big[a_m^{k,t}-a_m^{k,t-1}\Big]^+.
\end{align}
Therefore, if $F_{mn}^k\geq0, \forall m\neq n,k$ or $F_{mn}^k\leq0,
\forall m\neq n,k$, given (\ref{C:C2}), the sequence
$\{\textbf{a}_n^t\}$ contracts with the modulus $\rho(\textbf{T}^{\max})<1$ under the norm
$\max_{n\in\mathcal{N}}\frac{\sum_{k}[x_n^{k}]^+}{\bar{w}_n}$ and the convergence follows readily.
$\blacksquare$

\section{Proof of Theorem \ref{th:th2}}
Let $\|\cdot\|^\textbf{w}_2$ denote the weighted Euclidean norm with
weights $\textbf{w}=[w_1 \ldots w_K]^T$, i.e.
$\|\textbf{x}\|^\textbf{w}_2\triangleq(\sum_iw_i|x_i|^2)^{1/2}$
\cite{Matrix}. Define the simplex
\begin{equation}
\label{eq:eqnsim}
\mathcal{S}\triangleq\Bigl\{\mathbf{x}\in\mathcal{R}^K:
\frac{1}{K}\sum_{k=1}^Kx_k=1, x_k^{\min}\leq x_k\leq x_k^{\max}, \ \forall
k=1,2,\ldots,K\Bigr\}.
\end{equation}in which $\sum_kx_k^{\max}\geq1$.
The following lemma is needed to prove Theorem \ref{th:th2}.
\begin{lemma}\label{lm:lm2}
The projection with respect to the weighted Euclidean norm with
weights $\textbf{w}$, of the $K$-dimensional real
vector $-\mathbf{x}_0\triangleq-[x_{0,1},\ldots,x_{0,K}]^T$ onto the
simplex $\mathcal{S}$ defined in (\ref{eq:eqnsim}), denoted by
$[-\mathbf{x}_0]_{\mathcal{S}}^\textbf{w}$, is the
optimal solution to the following convex optimization problem:
\begin{equation}
[-\mathbf{x}_0]_{\mathcal{S}}^\textbf{w}\triangleq
\arg\min_{\mathbf{x}\in
\mathcal{S}}\bigl\|\mathbf{x}-(-\mathbf{x}_0)\bigr\|^\textbf{w}_2
\end{equation}
and takes the following form:
\begin{equation}
x_k^*=\Bigl[\frac{\lambda}{w_k}-x_{0,k}\Bigr]_{x_k^{\min}}^{x_k^{\max}}, k=1,\ldots,K
\end{equation}
where $\lambda>0$ is chosen in order to satisfy the constraint
$\frac{1}{K}\sum_{k=1}^Kx^*_k=1$.
\end{lemma}

\textit{Proof of Lemma \ref{lm:lm2}:} See Corollary 2 in
\cite{Scutari}. $\blacksquare$

For $h_n^{k}(\cdot)$ defined in (\ref{eq:eqn14}), user $n$ updates its
action according to
\begin{equation}
\label{eq:eqnA4}
a_n^{*k}=l_n^k(\mathbf{a}_{-n},\lambda^*)=\Big[\big(\frac{1}{F_{nn}^k}\big)^{1+\frac{1}{\theta}}\cdot(\lambda^*)^{\frac{1}{\theta}}
-\frac{\alpha_n^k}{F_{nn}^k}-\sum_{m\neq
n}F_{mn}^ka_m^k\Big]_{a_{n,k}^{\min}}^{a_{n,k}^{\max}}.
\end{equation}
and $\lambda^*$ is chosen to satisfy $\sum_{k=1}^Ka_n^{*k}=M_n$.
Define the vector update operator as
$[\textrm{BR}(\mathbf{a}_{-n})]_k\triangleq a_n^{*k}$ and the
coupling vector as
\begin{equation}
\label{eq:eqnA5}
[\textbf{C}_n(\mathbf{a}_{-n})]_k\triangleq\frac{\alpha_n^k}{F_{nn}^k}+\sum_{m\neq
n}F_{mn}^ka_m^k
\end{equation}
with $k \in \{1,\ldots,K\}$. We also define
\begin{equation}
\label{eq:eqnA6}
\textbf{F}'_{mn}\triangleq\textrm{diag}\Bigl(F_{mn}^1,F_{mn}^2,\ldots,F_{mn}^K\Bigr)
\end{equation}
and
\begin{equation}
\label{eq:eqnA7}
{\boldsymbol\alpha}'_{n}\triangleq\Bigl[\frac{\alpha_{n}^1}{F_{nn}^1},\frac{\alpha_{n}^2}{F_{nn}^2},\ldots,\frac{\alpha_{n}^K}{F_{nn}^K}\Bigr]^T.
\end{equation}
Therefore, the coupling vector can be alternatively rewritten as
\begin{equation}
\label{eq:eqnA8}
\textbf{C}_n(\mathbf{a}_{-n})={\boldsymbol\alpha}'_n+\sum_{m\neq
n}\textbf{F}'_{mn}\mathbf{a}_m.
\end{equation}
Define a weight matrix $\textbf{W}=[\textbf{w}_1 \ldots
\textbf{w}_N]$ in which the element $[\textbf{W}]_{kn}$ is chosen
according to
\begin{equation}
\label{eq:eqnA9}
[\textbf{W}]_{kn}=[\textbf{w}_n]_k=\frac{(F_{nn}^k)^{1+\frac{1}{\alpha}}}{\sum_{k=1}^K(F_{nn}^k)^{1+\frac{1}{\alpha}}}.
\end{equation}
By Lemma \ref{lm:lm2}, we know that the vector update operator
$\textrm{BR}_n(\mathbf{a}_{-n})$ in (\ref{eq:eqn15}) can be
interpreted as the projection of the coupling vector
$-\textbf{C}_n(\mathbf{a}_{-n})$ onto user $n$'s action set
$\mathcal{A}_n$ with respect to $\|\cdot\|^{\textbf{w}_n}_2$, i.e.
\begin{equation}
\label{eq:eqnA10}
\textrm{BR}_n(\mathbf{a}_{-n})=\bigl[-\textbf{C}_n(\mathbf{a}_{-n})\bigr]_{\mathcal{A}_n}^{\textbf{w}_n}.
\end{equation}
Given any $\mathbf{a}^{(1)},\mathbf{a}^{(2)}\in \mathcal{A}$, we
define respectively, for each user $n$, the weighted Euclidean
distances between these two vectors and their projected vectors
using (\ref{eq:eqnA10}) as
$e_n=\bigl\|\mathbf{a}_n^{(2)}-\mathbf{a}_n^{(1)}\bigr\|^{\textbf{w}_n}_2$
and
$e_{\textrm{BR}_n}=\bigl\|\textrm{BR}_n(\mathbf{a}_{-n}^{(1)})-\textrm{BR}_n(\mathbf{a}_{-n}^{(2)})\bigr\|^{\textbf{w}_n}_2$.
Again, we first prove the convergence of the parallel update case in
(\ref{eq:eqn12}). We have $\forall n\in \mathcal{N}$,
\begin{align}
&e_{\textrm{BR}_n}=\Bigl\|\bigl[-\textbf{C}_n(\mathbf{a}_{-n}^{(1)})\bigr]_{\mathcal{A}_n}^{\textbf{w}_n}-\bigl[-\textbf{C}_n(\mathbf{a}_{-n}^{(2)})\bigr]_{\mathcal{A}_n}^{\textbf{w}_n}\Bigr\|^{\textbf{w}_n}_2\nonumber\\
\label{eq:eq_tmp1}
&\leq\Bigl\|\textbf{C}_n(\mathbf{a}_{-n}^{(2)})-\textbf{C}_n(\mathbf{a}_{-n}^{(1)})\Bigr\|^{\textbf{w}_n}_2\\
&=\Bigl\|\sum_{m\neq
n}\textbf{F}'_{mn}\mathbf{a}_m^{(2)}-\sum_{m\neq
n}\textbf{F}'_{mn}\mathbf{a}_m^{(1)}\Bigr\|^{\textbf{w}_n}_2=\Bigl\|\sum_{m\neq
n}\textbf{F}'_{mn}\bigl(\mathbf{a}_m^{(2)}-\mathbf{a}_m^{(1)}\bigr)\Bigr\|^{\textbf{w}_n}_2\\
\label{eq:eq_tmp2} &\leq\sum_{m\neq
n}\Bigl\|\textbf{F}'_{mn}\bigl(\mathbf{a}_m^{(2)}-\mathbf{a}_m^{(1)}\bigr)\Bigr\|^{\textbf{w}_n}_2=\sum_{m\neq
n}\sqrt{\sum_{k=1}^K[\textbf{w}_n]_k\bigl([\textbf{F}'_{mn}]_{kk}\bigr)^2\Bigl(a_m^{(2)k}-a_m^{(1)k}\Bigr)^2}\\
&=\sum_{m\neq
n}\sqrt{\sum_{k=1}^K[\textbf{w}_m]_k\Bigl([\textbf{F}'_{mn}]_{kk}\frac{[\textbf{w}_n]_k}{[\textbf{w}_m]_k}\Bigr)^2\Bigl(a_m^{(2)k}-a_m^{(1)k}\Bigr)^2}\\
&\leq\sum_{m\neq
n}\max_k\Bigl(\big|[\textbf{F}'_{mn}]_{kk}\big|\cdot\frac{[\textbf{w}_n]_k}{[\textbf{w}_m]_k}\Bigr)\sqrt{\sum_{k=1}^K[\textbf{w}_m]_k\Bigl(a_m^{(2)k}-a_m^{(1)k}\Bigr)^2}\\
&=\sum_{m\neq
n}\max_k\Bigl(\big|[\textbf{F}'_{mn}]_{kk}\big|\cdot\frac{[\textbf{w}_n]_k}{[\textbf{w}_m]_k}\Bigr)\bigl\|\mathbf{a}_m^{(2)}-\mathbf{a}_m^{(1)}\bigr\|^{\textbf{w}_m}_2\\
\label{eq:eq_tmp3} &=\sum_{m\neq n}[\textbf{S}^{\max}]_{mn}e_m,
\end{align}
where (\ref{eq:eq_tmp1}) follows from the non-expansion property of
the projector $[\cdot]_{\mathcal{A}_n}^{\textbf{w}_n}$ in the norm
$\|\cdot\|^{\textbf{w}_n}_2$ (See Proposition 3.2(c) in
\cite{Bertsekas}), (\ref{eq:eq_tmp2}) follows from the triangle
inequality \cite{Matrix}, and $\textbf{S}^{\max}$ in
(\ref{eq:eq_tmp3}) is defined according to (\ref{eq:eqn16}).

The rest of the proof is similar as the proof after equation (\ref{eq:g5})
in Appendix A. Details are omitted due to space limitations. $\blacksquare$

\section{Proof of Theorem \ref{th:th3}}
The beginning part of the proof is the same as the proof of Theorem
\ref{th:th1}. For any user $n$ with general $f_n^k(\cdot)$, the
inequalities after (\ref{eq:g1}) become
\begin{align}
&\sum_{k=1}^K[a_n^{k,t+1}-a_n^{k,t}]^+\leq\max\{p^{n,t}(\lambda_n^{t}),q^{n,t}(\lambda_n^{t})\}\nonumber\\
\label{eq:gg1}&=\max\Bigg\{\sum_{k=1}^K\Big[f_n^k(\mathbf{a}_{-n}^t)-f_n^k(\mathbf{a}_{-n}^{t-1})\Big]^+,\sum_{k=1}^K\Big[f_n^k(\mathbf{a}_{-n}^t)-f_n^k(\mathbf{a}_{-n}^{t-1})\Big]^-\Bigg\}\\
\label{eq:gg2}&=\max\Bigg\{\sum_{k=1}^K\Big[\sum_{m \neq n}\sum_{k'=1}^K\frac{\partial f_n^k}{\partial a_m^{k'}}(\xi^{t}_{-n})(a_m^{k',t}-a_m^{k',t-1})\Big]^+,\sum_{k=1}^K\Big[\sum_{m \neq n}\sum_{k'=1}^K\frac{\partial f_n^k}{\partial a_m^{k'}}(\xi^{t}_{-n})(a_m^{k',t}-a_m^{k',t-1})\Big]^-\Bigg\}\\
\label{eq:gg3}&\leq\max\Bigg\{\sum_{k=1}^K\sum_{m \neq n}\sum_{k'=1}^K\Big[\frac{\partial f_n^k}{\partial a_m^{k'}}(\xi^{t}_{-n})(a_m^{k',t}-a_m^{k',t-1})\Big]^+,\sum_{k=1}^K\sum_{m \neq n}\sum_{k'=1}^K\Big[\frac{\partial f_n^k}{\partial a_m^{k'}}(\xi^{t}_{-n})(a_m^{k',t}-a_m^{k',t-1})\Big]^-\Bigg\}\\
\label{eq:gg4}&=\max\Bigg\{\sum_{m \neq n}\sum_{k'=1}^K\sum_{k=1}^K\Big[\frac{\partial f_n^k}{\partial a_m^{k'}}(\xi^{t}_{-n})(a_m^{k',t}-a_m^{k',t-1})\Big]^+,\sum_{m \neq n}\sum_{k'=1}^K\sum_{k=1}^K\Big[\frac{\partial f_n^k}{\partial a_m^{k'}}(\xi^{t}_{-n})(a_m^{k',t}-a_m^{k',t-1})\Big]^-\Bigg\}\\
\label{eq:gg5}&\leq\sum_{m \neq n}\Biggl\{\max_{k'}\sum_{k=1}^K\Big|\frac{\partial f_n^k}{\partial a_m^{k'}}(\xi^{t}_{-n})\Big|\Biggr\}\cdot\Bigg\{\sum_{k'=1}^K\Big[(a_m^{k',t}-a_m^{k',t-1})\Big]^++\sum_{k'=1}^K\Big[(a_m^{k',t}-a_m^{k',t-1})\Big]^-\Bigg\}\\
\label{eq:gg6}&=\sum_{m \neq
n}2\cdot\Biggl\{\max_{k'}\sum_{k=1}^K\Big|\frac{\partial
f_n^k}{\partial
a_m^{k'}}(\xi^{t}_{-n})\Big|\Biggr\}\cdot\sum_{k'=1}^K\Big[(a_m^{k',t}-a_m^{k',t-1})\Big]^+
\end{align}
where (\ref{eq:gg1}) follows from the definition of $p^{n,t}$ and
$q^{n,t}$ and the expression of $a_n^{k,t}$ and
$B_n^k(\mathbf{a}_{-n},\lambda)$ in (\ref{eq:eqn12}) and
(\ref{eq:newBR}), (\ref{eq:gg2}) follows from the mean value theorem
for vector-valued functions with
$\xi^{t}=\alpha\mathbf{a}^t+(1-\alpha)\mathbf{a}^{t-1}$ and
$\alpha\in[0,1]$. By (\ref{C:C4}), it is straightforward to show
that the iteration is a contraction by following the same arguments
in Appendix A. The rest of the proof is omitted. $\blacksquare$

\section{Proof of Theorem \ref{th:th4}}
The gradient play algorithm in (\ref{eq:eqn31}) is in fact a
gradient projection algorithm with constant stepsize $\kappa$. In
order to establish its convergence, we first need to prove that the
gradient of the objective in (\ref{eq:eqn25}) is Lipschitz
continuous, with a Lipschitz constant given by $L>0$, i.e.
\begin{equation}
\label{eq:eqnA12} \Big\|\bigtriangledown\big(\sum_{n=1}^N
u_n(\mathbf{x})\big)-\bigtriangledown\big(\sum_{n=1}^N
u_n(\mathbf{y})\big)\Big\|\leq L\big\|\mathbf{x}-\mathbf{y}\big\|, \
\forall \mathbf{x},\mathbf{y}\in\mathcal{A}.
\end{equation}
It is known that it has the property of Lipschitz continuity if it
has a Hessian bounded in the Euclidean norm.

The Hessian matrix $\textbf{H}$ of $\sum_{n=1}^N u_n(\mathbf{a})$
can be decomposed into two matrices:
$\textbf{H}=\textbf{H}_1+\textbf{H}_2$, in which the elements of
matrix $\textbf{H}_1$ are
\begin{equation}
\frac{\partial^2\Big[\sum\limits_{n=1}^N\sum\limits_{k=1}^Kh_n^k\big(a_{n}^k+\sum\limits_{m\neq
n}F_{mn}^ka_m^k\big)\Big]}{\partial a_m^k\partial a_l^j}=\left\{
\begin{array}{cl}
\sum\limits_{n=1}^N\frac{\partial^2 h_n^{k}}{\partial^2
x}\big(a_{n}^k+\sum\limits_{m\neq n}F_{mn}^ka_m^k\big)F_{mn}^kF_{ln}^k,& \text{if $k=j$}\\
0,& \text{otherwise}.
\end{array} \right.
%\frac{\partial^2\Big[\sum_{n=1}^N\sum_{k=1}^Kh_n^k\big(a_{n}^k+f_n^k(\mathbf{a}_{-n})\big)\Big]}{\partial
%a_m^k\partial a_l^j}= \sum_{n=1}^N\sum_{k=1}^K\frac{\partial^2
%h_n^{k}}{\partial^2
%x}\big(a_{n}^k+f_n^k(\mathbf{a}_{-n})\big)\cdot\frac{\partial^2\big(a_{n}^k+f_n^k(\mathbf{a}_{-n})\big)}{\partial
%a_m^k\partial a_l^j}
\end{equation}
with $F_{nn}^k=1$ and the elements of matrix $\textbf{H}_2$ are
\begin{equation}
-\frac{\partial^2\Big[\sum\limits_{n=1}^N\sum\limits_{k=1}^Kg_n^k(\mathbf{a}_{-n})\Big]}{\partial
a_m^k\partial
a_l^j}=-\sum_{n=1}^N\sum_{k=1}^K\frac{\partial^2g_n^k(\mathbf{a}_{-n})}{\partial
a_m^k\partial a_l^j}.
%=\left\{
%\begin{array}{cl}
%\sum_{n\neq m,l}\frac{\partial^2 g_n^k(\mathbf{a}_{-n})}{\partial
%a_m^k\partial
%a_l^k},& \text{if $k=j$}\\
%0,& \text{otherwise}.
%\end{array} \right.
\end{equation}
Recall that $g_n^k(\cdot)$ is Lipschitz continuous and it satisfies
\begin{displaymath}
\Big\|\bigtriangledown g_n^k(\mathbf{x})-\bigtriangledown
g_n^k(\mathbf{y})\Big\|\leq L'\big\|\mathbf{x}-\mathbf{y}\big\|, \
\forall n,k,\mathbf{x},\mathbf{y}\in\mathcal{A}_{-n}.
\end{displaymath}
Consequently, we have $\|\textbf{H}_2\|_2\leq NKL'$. As a result, we
can estimate the Lipschitz constant $L$ using the following
inequalities
\begin{equation}
\label{eq:Lup}
\|\textbf{H}\|_2\leq\|\textbf{H}_1\|_2+\|\textbf{H}_2\|_2\leq\sqrt{\|\textbf{H}_1\|_1\|\textbf{H}_1\|_\infty}+NKL'\leq
\sup_{x,n,k}\Big|\frac{\partial^2 h_n^{k}}{\partial^2
x}\Big|\cdot\max_{k,l}\sum_{m=1}^N\sum_{n=1}^N|F_{mn}^kF_{ln}^k|+NKL'.
\end{equation}
We can choose the RHS of (\ref{eq:Lup}) as the Lipschitz constant
$L$ . By Proposition 3.4 in \cite{Bertsekas}, we know that if
$0<\kappa<2/L$, the sequence $\mathbf{a}^t$ generated by the
gradient projection algorithm in (\ref{eq:eqn32}) and
(\ref{eq:eqn33}) converges to a limiting point at which the KKT
conditions in (\ref{eq:eqn26})-(\ref{eq:eqn28}) are satisfied.
$\blacksquare$

\section{Proof of Theorem \ref{th:th5}}
We know from the proof of Theorem \ref{th:th4} that, under Condition
(\ref{C:C7}), $\sum_{n=1}^N u_n(\mathbf{a})$ is Lipschitz continuous
and the inequality in (\ref{eq:eqnA12}) holds. Recall that
$\sum_{n=1}^N u_n(\mathbf{x})$ is continuously differentiable.
Therefore, by the descent lemma \cite{Bertsekas}, we have
\begin{equation}
\label{eq:eqnA13} \sum_{n=1}^N u_n(\mathbf{x})\geq\sum_{n=1}^N
u_n(\mathbf{y})+(\mathbf{x}-\mathbf{y})^T\cdot\bigtriangledown\big(\sum_{n=1}^N
u_n(\mathbf{y})\big)-\frac{L}{2}\big\|\mathbf{x}-\mathbf{y}\big\|_2^2,
\ \forall \mathbf{x},\mathbf{y}\in\mathcal{A}.
\end{equation}
Therefore, in order to prove $\sum_{n=1}^N u_n(\mathbf{a}^t)\geq
\sum_{n=1}^N u_n(\mathbf{a}^{t-1})$, we only need to show that
\begin{equation}
\label{eq:eqnA14}
(\mathbf{a}^t-\mathbf{a}^{t-1})^T\cdot\bigtriangledown\big(\sum_{n=1}^N
u_n(\mathbf{a}^{t-1})\big)\geq\frac{L}{2}\big\|\mathbf{a}^t-\mathbf{a}^{t-1}\big\|_2^2
\end{equation} for sufficiently small $\kappa$. Substituting (\ref{eq:eqn35})
into (\ref{eq:eqnA14}), we can see that it is equivalent to
\begin{equation}
\label{eq:eqnA14.5}\sum_{n=1}^N\sum_{k=1}^K\bigl(B_n^{'k}(\textbf{a}_{-n}^{t-1})-a_n^{k,t-1}\bigr)\cdot\frac{\partial\sum_{n=1}^N
u_n(\mathbf{a}^{t-1})}{\partial
a_n^{k,t-1}}\geq\kappa\cdot\frac{L}{2}\cdot\sum_{n=1}^N\sum_{k=1}^K\bigl(B_n^{'k}(\textbf{a}_{-n}^{t-1})-a_n^{k,t-1}\bigr)^2.
\end{equation}

By equation (\ref{eq:eqn34}), we have
\begin{equation}
\label{eq:eqnA15} B_n^{'k}(\textbf{a}_{-n}^{t-1})-a_n^{k,t-1}=
\Big\{\frac{\partial h^{k}_n}{\partial
x}\Big\}^{-1}\big(\lambda_n+\nu^{k}_n-\nu^{'k}_n+\sum_{m\neq
n}\pi_{mn}^{k,t-1}\big)-\sum_{m\neq
n}^NF_{mn}^ka_m^{k,t-1}-a_n^{k,t-1}
\end{equation}
and
\begin{equation}
\label{eq:eqnA16} \frac{\partial\sum_{n=1}^N
u_n(\mathbf{a}^{t-1})}{\partial a_n^{k,t-1}}=\frac{\partial
h^{k}_n}{\partial x}\big(a_n^{k,t-1}+\sum_{m\neq
n}^NF_{mn}^ka_m^{k,t-1}\big)-\sum_{m\neq n}\pi_{mn}^{k,t-1}.
\end{equation}
By the mean value theorem, there exists $\xi_n^k\in\mathcal{R}$ such
that
\begin{align}
&\frac{\partial h^{k}_n}{\partial x}\big(a_n^{k,t-1}+\sum_{m\neq
n}^NF_{mn}^ka_m^{k,t-1}\big)-\sum_{m\neq n}\pi_{mn}^{k,t-1}\nonumber\\
&=\frac{\partial h^{k}_n}{\partial x}\big(a_n^{k,t-1}+\sum_{m\neq
n}^NF_{mn}^ka_m^{k,t-1}\big)-\Big(\lambda_n+\nu^{k}_n-\nu^{'k}_n+\sum_{m\neq
n}\pi_{mn}^{k,t-1}\Big)+\Big(\lambda_n+\nu^{k}_n-\nu^{'k}_n\Big)\nonumber\\
&=\frac{\partial^2 h^{k}_n}{\partial^2
x}(\xi_n^k)\cdot\Bigg\{a_n^{k,t-1}+\sum_{m\neq
n}^NF_{mn}^ka_m^{k,t-1}-\Big\{\frac{\partial h^{k}_n}{\partial
x}\Big\}^{-1}\Big(\lambda_n+\nu^{k}_n-\nu^{'k}_n+\sum_{m\neq
n}\pi_{mn}^{k,t-1}\Big)\Bigg\}+\lambda_n+\nu^{k}_n-\nu^{'k}_n.\nonumber
\end{align}
Multiplying (\ref{eq:eqnA15}) and (\ref{eq:eqnA16}) leads to
\begin{align}
\label{eq:eqnA17}&\sum_{n=1}^N\sum_{k=1}^K\bigl(B_n^{'k}(\textbf{a}_{-n}^{t-1})-a_n^{k,t-1}\bigr)\cdot\frac{\partial\sum_{n=1}^N
u_n(\mathbf{a}^{t-1})}{\partial a_n^{k,t-1}}\nonumber\\
&=-\sum_{n=1}^N\sum_{k=1}^K\frac{\partial^2 h^{k}_n}{\partial^2
x}(\xi_n^k)\cdot\Bigg\{\Big\{\frac{\partial h^{k}_n}{\partial
x}\Big\}^{-1}\Big(\lambda_n+\nu^{k}_n-\nu^{'k}_n+\sum_{m\neq
n}\pi_{mn}^{k,t-1}\Big)-a_n^{k,t-1}-\sum_{m\neq
n}^NF_{mn}^ka_m^{k,t-1}\Bigg\}^2\nonumber\\
&+\sum_{n=1}^N\sum_{k=1}^K\bigl(B_n^{'k}(\textbf{a}_{-n}^{t-1})-a_n^{k,t-1}\bigr)\cdot(\lambda_n+\nu^{k}_n-\nu^{'k}_n)
\end{align}
In the following, we differentiate two cases in which the Lagrange
multipliers $\lambda_n,\nu^{k}_n,\nu^{'k}_n$ take different values.

First of all, if $\lambda_n=\nu^{k}_n=\nu^{'k}_n=0$ for all $k,n$,
equation (\ref{eq:eqnA17}) can be simplified as
\begin{align}
&\sum_{n=1}^N\sum_{k=1}^K\bigl(B_n^{'k}(\textbf{a}_{-n}^{t-1})-a_n^{k,t-1}\bigr)\cdot\frac{\partial\sum_{n=1}^N
u_n(\mathbf{a}^{t-1})}{\partial a_n^{k,t-1}}\nonumber\\
&=-\sum_{n=1}^N\sum_{k=1}^K\frac{\partial^2 h^{k}_n}{\partial^2
x}(\xi_n^k)\cdot\Bigg\{\Big\{\frac{\partial h^{k}_n}{\partial
x}\Big\}^{-1}\Big(\lambda_n+\nu^{k}_n-\nu^{'k}_n+\sum_{m\neq
n}\pi_{mn}^{k,t-1}\Big)-a_n^{k,t-1}-\sum_{m\neq
n}^NF_{mn}^ka_m^{k,t-1}\Bigg\}^2.
\end{align}

On the other hand, if $\lambda_n>0$, $\nu^{k}_n>0$, or
$\nu^{'k}_n>0$ for some $k,n$. Due to complementary slackness in
(\ref{eq:eqn27}) and (\ref{eq:eqn28}), We know that
\begin{align}
&\lambda_n>0 \Rightarrow
\sum_{k=1}^KB_n^{'k}(\textbf{a}_{-n}^{t-1})=M_n\geq\sum_{k=1}^Ka_n^{k,t-1},\nonumber\\
&\nu^{k}_n>0 \Rightarrow
B_n^{'k}(\textbf{a}_{-n}^{t-1})=a_{n,k}^{\max}\geq a_n^{k,t-1}, \nonumber\\
&\nu^{'k}_n>0, \Rightarrow
B_n^{'k}(\textbf{a}_{-n}^{t-1})=a_{n,k}^{\min}\leq a_n^{k,t-1}.
\nonumber
\end{align}
As a result, the last term in (\ref{eq:eqnA17}) satisfy
\begin{align}
&\sum_{n=1}^N\sum_{k=1}^K\bigl(B_n^{'k}(\textbf{a}_{-n}^{t-1})-a_n^{k,t-1}\bigr)\cdot(\lambda_n+\nu^{k}_n-\nu^{'k}_n)=\sum_{n=1}^N\lambda_n\sum_{k=1}^K(B_n^{'k}(\textbf{a}_{-n}^{t-1})-a_n^{k,t-1}\bigr)+\nonumber\\
&\sum_{n=1}^N\sum_{k=1}^K\nu^{k}_n\bigl(B_n^{'k}(\textbf{a}_{-n}^{t-1})-a_n^{k,t-1}\bigr)
+\sum_{n=1}^N\sum_{k=1}^K\nu^{'k}_n\bigl(a_n^{k,t-1}-B_n^{'k}(\textbf{a}_{-n}^{t-1})\bigr)\geq0.
\end{align}
Therefore, in both cases, the following inequality holds
\begin{align}
&\sum_{n=1}^N\sum_{k=1}^K\bigl(B_n^{'k}(\textbf{a}_{-n}^{t-1})-a_n^{k,t-1}\bigr)\cdot\frac{\partial\sum_{n=1}^N
u_n(\mathbf{a}^{t-1})}{\partial a_n^{k,t-1}}\nonumber\\
&\geq-\sum_{n=1}^N\sum_{k=1}^K\sup_{x}\frac{\partial^2
h^{k}_n(x)}{\partial^2 x}\cdot\Bigg\{\Big\{\frac{\partial
h^{k}_n}{\partial
x}\Big\}^{-1}\Big(\lambda_n+\nu^{k}_n-\nu^{'k}_n+\sum_{m\neq
n}\pi_{mn}^{k,t-1}\Big)-a_n^{k,t-1}-\sum_{m\neq
n}^NF_{mn}^ka_m^{k,t-1}\Bigg\}^2\nonumber\\
&=-\sum_{n=1}^N\sum_{k=1}^K\sup_{x}\frac{\partial^2
h^{k}_n(x)}{\partial^2
x}\cdot\bigl(B_n^{'k}(\textbf{a}_{-n}^{t-1})-a_n^{k,t-1}\bigr)^2.
\end{align}
Finally, we can conclude that the inequality in (\ref{eq:eqnA14.5})
holds for
$\kappa\leq\frac{2}{L}\cdot(-\max_{n,k}\sup_{x}\frac{\partial^2
h^{k}_n(x)}{\partial^2 x})$. Recall that Jacobi update requires
$\kappa\in(0,1]$. The stepsize $\kappa$ can be eventually chosen as
$0<\kappa\leq\min\{\frac{2}{L}\cdot(-\max_{n,k}\sup_{x}\frac{\partial^2
h^{k}_n(x)}{\partial^2 x}),1\}$ $\blacksquare$

\section{Upper bound of $\rho(\textbf{S}^k)$}
Denote $\textbf{\textrm{1}}^T=[1\ 1\cdots 1]^T$. If we fix
$1-r_{m0}^k$ for $\forall m,k$, we have
$\textbf{\textrm{1}}^T\textrm{R}^k=(1-r_{m0}^k)\textbf{\textrm{1}}^T$.
Note that
$\Upsilon^k=(\textrm{I}-\textrm{R}^k)^{-1}=\textrm{I}+\sum_{i=1}^{\infty}(\textrm{R}^k)^i$.
We have
$\textbf{\textrm{1}}^T\Upsilon^k=\textbf{\textrm{1}}^T\big(\textrm{I}+\sum_{i=1}^{\infty}(\textrm{R}^k)^i\big)
=\textbf{\textrm{1}}^T+(1-r_{m0}^k)\textbf{\textrm{1}}^T\Upsilon^k$
and
$\textbf{\textrm{1}}^T\Upsilon^k=\frac{1}{1-r_{m0}^k}\textbf{\textrm{1}}^T$.
Therefore, $|\Upsilon^k|_1=\frac{1}{r_{m0}^k}$. Since
$F^k_{mn}=\frac{[\Upsilon^k]_{nm}}{[\Upsilon^k]_{nn}}$ and
$\Upsilon^k=\textrm{I}+\sum_{i=1}^{\infty}(\textrm{R}^k)^i$, we know
$[\Upsilon^k]_{nn}\geq1$ for $\forall n$. Denote a diagonal matrix
$\textrm{diag}(\Upsilon^k)$ with the entries of $\Upsilon^k$ on the
diagonal. Recall that $[\textbf{S}^k]_{mn}=F^k_{mn}$ for $m\neq n$,
and $[\textbf{S}^k]_{nn}=0$ for $n\in \mathcal{N}$. We can conclude
that
$\rho(\textbf{S}^k)\leq|\textbf{S}^k|_{\infty}\leq|(\Upsilon^k)^T-\textrm{diag}(\Upsilon^k)|_{\infty}\leq|(\Upsilon^k)^T|_{\infty}-1=|\Upsilon^k|_{1}-1=\frac{1}{r_{m0}^k}-1$.

\ifCLASSOPTIONcaptionsoff
  \newpage
\fi

\newpage

\end{document}